\documentclass[aps, prb, twocolumn, superscriptaddress, english]{revtex4-2}

\usepackage{graphicx}
\usepackage{cancel}
\usepackage{enumerate}
\usepackage{hyperref}
\hypersetup{
colorlinks=true,
citecolor=blue,
linkcolor=blue,
urlcolor=blue,
}
\usepackage{textcomp}
\usepackage{amsmath}
\usepackage{amssymb}
\usepackage{pgf}
\usepackage{soul}
\usepackage{subfigure}
\usepackage[english]{babel}
\usepackage[normalem]{ulem}

\def\bra#1{\langle#1|}
\def\ket#1{|#1\rangle}

\def\ee{\mathrm{e}}

\newcommand{\Up}{{\uparrow}}
\newcommand{\Dn}{{\downarrow}}
\newcommand{\Upt}{{{\sf u}}}
\newcommand{\Dnt}{{{\sf d}}}

\graphicspath{{./}{./Figure/}}

\begin{document}
\title{The poor man's Majorana tetron}

\author{Maximilian Nitsch}
\thanks{These authors contributed equally to this work.}
\affiliation{Division of Solid State Physics and NanoLund, Lund University, S-22100 Lund, Sweden}

\author{Lorenzo Maffi}
\thanks{These authors contributed equally to this work.}
\affiliation{Dipartimento di Fisica e Astronomia “G. Galilei”,
Universit\`a degli Studi di Padova, I-35131 Padova, Italy}
\affiliation{Istituto Nazionale di Fisica Nucleare (INFN), Sezione di Padova, I-35131 Padova, Italy}

\author{Virgil V. Baran}
\affiliation{Faculty of Physics, University of Bucharest, 405 Atomi\c stilor, RO-077125, Bucharest-M\u agurele, Romania}
\affiliation{``Horia Hulubei" National Institute of Physics and Nuclear Engineering, 30 Reactorului, RO-077125, Bucharest-M\u agurele, Romania}
\affiliation{Center for Quantum Devices, Niels Bohr Institute, Copenhagen University, 2100 Copenhagen, Denmark}

\author{Rubén Seoane Souto}
\affiliation{Instituto de Ciencia de Materiales de Madrid (ICMM),
Consejo Superior de Investigaciones Cient\'ificas (CSIC),
Sor Juana In\'es de la Cruz 3, 28049 Madrid, Spain}

\author{Jens Paaske}
\affiliation{Center for Quantum Devices, Niels Bohr Institute, Copenhagen University, 2100 Copenhagen, Denmark}

\author{Martin Leijnse}
\affiliation{Division of Solid State Physics and NanoLund, Lund University, S-22100 Lund, Sweden}

\author{Michele Burrello}
\affiliation{Center for Quantum Devices, Niels Bohr Institute, Copenhagen University, 2100 Copenhagen, Denmark}
\affiliation{Niels Bohr International Academy, Niels Bohr Institute, Copenhagen University, 2100 Copenhagen, Denmark}
\affiliation{Dipartimento di Fisica dell’Università di Pisa and INFN, Largo Pontecorvo 3, I-56127 Pisa, Italy}

\begin{abstract}
The Majorana tetron is a prototypical topological qubit stemming from the ground state degeneracy of a superconducting island hosting four Majorana modes. This degeneracy manifests as an effective non-local spin degree of freedom, whose most paradigmatic signature is the topological Kondo effect. Degeneracies of states with different fermionic parities characterize also minimal Kitaev chains which have lately emerged as a platform to realize and study unprotected versions of Majorana modes, dubbed poor man's Majorana modes. Here, we introduce the ``poor man's Majorana tetron'', comprising four quantum dots coupled via a floating superconducting island. Its charging energy yields non-trivial correlations among the dots, although, unlike a standard tetron, it is not directly determined by the fermionic parity of the Majorana modes. The poor man's tetron displays parameter regions with a two-fold degenerate ground state with odd fermionic parity, that gives rise to an effective Anderson impurity model when coupled to external leads. We show that this system can approach a regime featuring the topological Kondo effect under a suitable tuning of experimental parameters. Therefore, the poor man's tetron is a promising device to observe the non-locality of Majorana modes and their related fractional conductance.
\end{abstract}

\maketitle

\section{Introduction}\label{sec:intro}

The last decade witnessed a steady progress of the fabrication techniques of semiconductor-superconductor hybrid systems, driven, in particular, by the research on Majorana zero-energy modes and the engineering of artificial topological superconductors \cite{Prada2020,Flensberg2021}. These advances allowed for the combination of quantum dots and superconducting islands in the same controllable architecture, with nanowire devices coated by aluminum shells constituting the paradigmatic platform to engineer tunable superconducting devices. These systems take advantage of the interplay between the intrinsic spin-orbit coupling of the semiconductor nanowires and the superconducting pairing induced by proximity with aluminum. Additionally, they offer the possibility of creating superconducting subgap states that can be controlled through electrostatic gates, whose rich physics can be investigated via transport measurements.

These general features allowed for the realization of efficient Cooper pair splitters \cite{Wang2022,Wang2023} and minimal Kitaev chains \cite{Bordin2023,Dvir2023,Zatelli2023,bordin2024,haaf2024}, which constitute the basic platform to study the so-called poor man's Majorana modes \cite{Leijnse2012}. These are non-Abelian zero-energy modes that can be obtained through a fine-tuning of the parameters of the system \cite{Leijnse2012,Liu2022} and, for this reason, they are not protected against external perturbations, as in the case of a two-site Kitaev chain. Topological protection can be recovered by extending these systems towards longer quantum dot chains, thus progressively approaching the topological phase of the Kitaev model \cite{Sau2012,Fulga2013,Svensson2024}. Compared with the conventional microscopic realization of Majorana modes in nanowires \cite{Oreg2010,Lutchyn2010}, this mesoscopic approach offers additional control to overcome the limitations imposed by disorder. Therefore it constitutes a promising alternative for observing properties of Majorana modes that were so far elusive.

In minimal devices with two quantum dots coupled through an intermediate superconducting island via weak tunneling amplitudes, two main inter-dot interactions emerge. The first is the crossed Andreev reflection (CAR), corresponding to a process in which two electrons, one for each dot, couple to form a Cooper pair. The second is the elastic cotunneling of electrons from one dot to the other. Poor man's Majorana modes can be realized when the amplitudes of these processes match. Such condition can be achieved by mediating these processes through controllable Andreev bound states \cite{Bordin2023,Liu2022} in the intermediate superconductor. The onset of poor man's Majorana modes, however, does not necessarily require these additional states, and alternative techniques for their realization have been proposed based on the control of magnetic fluxes and superconducting phases \cite{Fulga2013,Samuelson2024,Svensson2024}.

The experimental investigations of minimal Kitaev chains \cite{Dvir2023,Zatelli2023,bordin2024,haaf2024} have been realized in setups with grounded superconducting islands, such that the charging energy of the electrons in the superconductors can be neglected \cite{SeoaneSouto2024}. The study of Majorana modes, however, demonstrated that devices with floating superconducting islands display additional transport effects that can be used for their characterization \cite{Higginbotham_NatPhys2015,Albrecht_Nature2016,vanHeck_PRB2016,Albrecht_PRL2017,Shen_NatComm2018}. Furthermore, electrostatic interactions constitute a key ingredient to engineer one of the most crucial building blocks for the design of Majorana platforms for quantum information processing: the tetron \cite{Karzig2017}. A tetron is a floating superconducting island, or Cooper-pair box, hosting four Majorana modes, and it represents one of the paradigmatic examples of topological qubit \cite{Terhal2012,Landau_PRL2016,Plugge2016,Plugge2017}.
It is therefore a fundamental question to assess whether systems combining minimal Kitaev chains and charging energy effects can be used to define possible qubit platforms and explore the non-local properties of Majorana modes.

\begin{figure}[t!]
\centering
\includegraphics[width=\columnwidth]{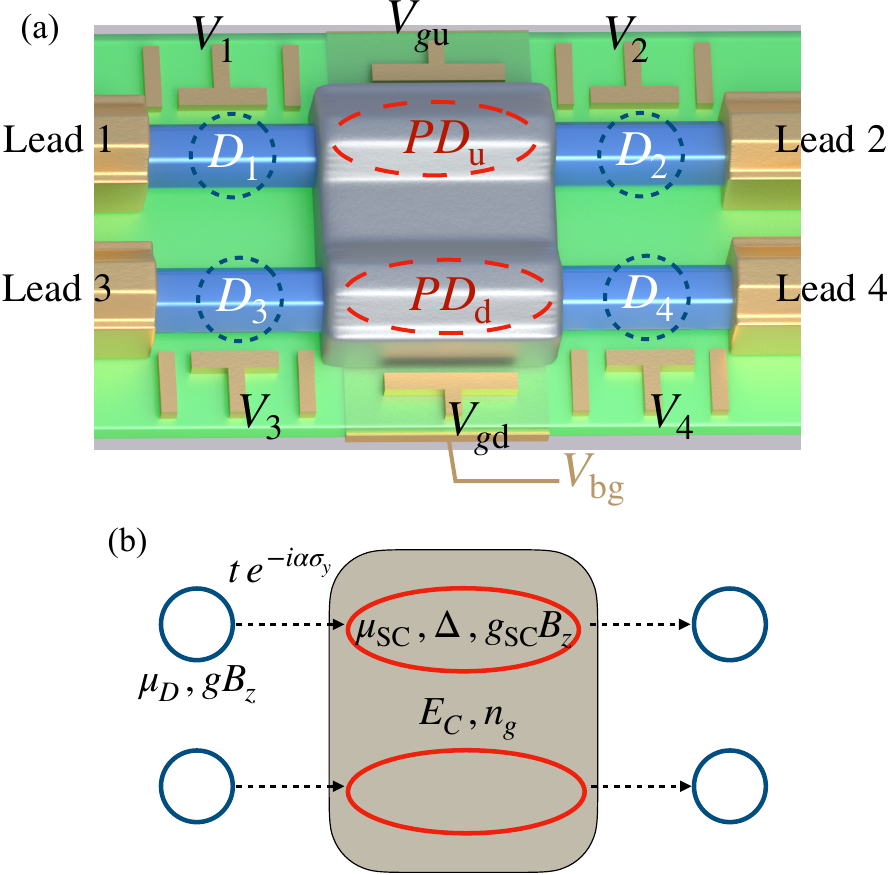}
\caption{(a) Sketch of the poor man's tetron: two semiconducting nanowires (blue) are connected by a shared SC floating island (silver); a set of electrostatic gates is used to define a quantum dot $D_i$ on each side of each nanowire (dashed circles). The superconducting segments of the nanowires are treated as proximitized dots $PD_{\Upt,\Dnt}$ (orange ellipses). Gate voltages are used to control the external dot energy levels $\mu_D$ ($V_i$ voltages), and the induced charge $n_g$ ($V_{g\tau}$ voltages). The voltage $V_{\rm bg}$ of a backgate is used to tune the potential $\mu_{\rm SC}$ in the proximitized dots. The device is then connected with four external leads (lateral gold contacts) to explore the emerging Kondo physics. (b) Schematic representation of the Hamiltonian terms of the poor man's tetron. The arrows indicate the sign of the spin-orbit coupling in the tunneling term.}
\label{fig:tetron-schematic}
\end{figure}

In this work we analyze to what extent the physics of tetrons can be recovered with poor man's Majorana modes. In particular we investigate the fate of these modes in devices with four quantum dots coupled with a single floating superconducting island (see Fig. \ref{fig:tetron-schematic}).  

It is known that intra-dot interactions can be beneficial for the formation of poor man's Majorana modes \cite{Tsintzis2022} and weak inter-dot interactions are not detrimental \cite{Samuelson2024,Brunetti2013}. 
Furthermore, in two-dot setups, Majorana sweet spots can be identified also in the presence of a sizeable charging energy in the intermediate superconducting island \cite{SoutoBaran}. We will show instead that the situation is different in devices with four dots. Here, in general, the charging energy of the superconductor destroys the poor man's Majorana modes. However, when suitably tuning the dot potentials, regimes with degenerate ground states can still be obtained. Therefore, we label such 4-dot systems {\it poor man's tetrons}. We envision that these devices can be fabricated by doubling the minimal Kitaev chain analyzed in Ref.~\cite{Dvir2023} with the use of double nanowires \cite{Vekris2022,Kanne2022} connected through a shared aluminum shell. Alternative devices fabricated in two-dimensional systems with gate-defined dots can be adopted as well to obtain analogous controllable platforms \cite{haaf2024}.

One of the key characteristics of the topologically protected tetron is the emergence of a non-local effective spin 1/2 degree of freedom that couples in a non-trivial way with external leads in multiterminal transport experiments.
In particular, the interaction of the tetron's Majorana modes with $M=3$ or $4$ leads is predicted to generate the so-called topological Kondo effect \cite{Beri2012,Beri2013,Altland2013}: in the low-temperature limit the tetron displays a fractional non-local conductance $(2/M)e^2/h$ corresponding to the onset of a non-Fermi liquid fixed point in the limit of strong coupling between the leads and the Majorana modes. These peculiar transport features, however, have never been observed so far: engineering a physical platform to study the topological Kondo effect is still considered a crucial milestone in the study of Majorana modes and, in general, of non-Abelian anyons. Indeed, recent theoretical works shed light on the relationship between the non-Fermi liquid fixed point in this kind of quantum impurity problems and the emergence of non-Abelian topological objects \cite{Papaj2019,Han2022,Lotem2022}.

In the following we investigate whether the correlated ground states that characterize a poor man's tetron can reproduce the exotic transport features identifying the topological Kondo effect, and whether other kinds of Kondo effects may emerge in this system.

In Sec. \ref{sec:model} we present the general model for a poor man's tetron composed by four tunable dots whose interactions are mediated by a single floating superconducting island. In Sec. \ref{sec:pert}, we analyze these devices in the perturbative regime with weak interdot tunneling and strong superconducting gap. 
We show in Sec. \ref{PMtetron} that the many-body states of the poor man's tetron display non-local correlations reminiscent of the topological tetron. In Sec. \ref{Kondo} we discuss how these correlations reflect in multi-terminal transport experiments, and we show that, upon a suitable tuning of the dot potentials and subgap state energy levels, it is possible to reach an effective topological Kondo point \cite{Beri2012,Beri2013,Altland2013}, characterized by fractional non-local conductance. Sec. \ref{sec:surrogate} analyzes the low-energy physics of poor man's tetron beyond perturbation theory through exact diagonalization. We will investigate, in particular, the extension in parameter space of the regime where the poor man's tetron behaves as a two-level system.

The appendices provide technical details on the perturbative analysis of the poor man's tetron and the definition of the related Kondo problems.

\section{The poor man's tetron}\label{sec:model}

Andreev bound states constitute a convenient tool to mediate the interactions between two quantum dots in a controllable way and realize poor man's Majorana modes.
In recent experiments \cite{Dvir2023}, this scenario has been realized by fabricating InSb nanowires which present a central segment covered by an aluminum superconductor and a pair of quantum dots, one on each side of the superconducting segment, defined in the nanowire through suitable electrostatic gates. Analogous results have been obtained also in a regime of strong hybridization between the dots and the central superconducting region, that leads to the formation of Yu-Shiba-Rusinov states mediating the interdot interactions \cite{Zatelli2023}.

In the limit of small quantum dots, the low-energy physics of each dot can be captured by a single energy-level, whose energy $\mu_D$ can be tuned via suitable electrostatic gates. Additionally, also the superconducting segment of the nanowire can be represented in its simplest approximation by a single-level central dot, subject to an s-wave superconducting pairing induced by proximity with the aluminum \cite{Liu2022,Tsintzis2022}.

The construction of the poor man's tetron (see Fig.~\ref{fig:tetron-schematic}) relies on doubling this system. We consider, in particular, two parallel nanowires that share the same floating superconducting shell \cite{Vekris2022,Kanne2022}. Each nanowire, labelled by $\tau=\Upt,\Dnt$, hosts two external single-level dots and an intermediate superconducting segment, characterized by a spin-1/2 doublet of Andreev subgap states. The Andreev states in each nanowire are tunnel-coupled with both quantum dots in the same wire; however, we assume that the Andreev states in different nanowires are independent, which corresponds to the situation in which the wires are separated and connected exclusively through the superconductor. As a consequence, the two nanowires interact only through the shared charging energy of the common superconducting shell, without any single-electron tunneling or crossed superconducting pairing between them. This scenario has been considered, for instance, in Ref.~\cite{Souto_PRB2022} in the context of multiterminal transport spectroscopy across hybrid semiconductor-superconductor double nanowires.

We decompose the Hamiltonian of the poor man's tetron in the form $H_{\rm sys}= H_{\rm sp} + H_{\rm c}$, where $H_{\rm sp}$ is the quadratic Hamiltonian describing the single-particle dynamics of the electrons in the two nanowires; $H_{\rm c}$ defines instead the charging energy of the superconducting island and may include also the Coulomb interactions of electrons within the same dot.

We approximate the single-particle Hamiltonian in the following way:
\begin{multline} \label{hamtot}
H_{\rm sp} = \sum_{\substack{a=L,R \\ s=\Up,\Dn \\ \tau=\Upt,\Dnt}} \mu_{\rm D} d^\dag_{as\tau} d_{as\tau} + gB_z\sum_{\substack{s,s'=\Up,\Dn \\ a=L,R \\ \tau=\Upt,\Dnt}} d^\dag_{as\tau} \sigma^z_{ss'} d_{as'\tau} \\ + g_{\rm SC}B_z\sum_{\tau,s,s'=\Up,\Dn} c^\dag_{s\tau} \sigma^z_{ss'} c_{s'\tau} +
\sum_{\substack{s=\Up,\Dn \\ \tau=\Upt,\Dnt}} \mu_{\rm SC} c^\dag_{s\tau}c_{s\tau}  \\ +\Delta \sum_{\tau=\Upt,\Dnt} \left(c^\dag_{\Up \tau}c^\dag_{\Dn \tau}\ee^{i\varphi} +{\rm H.c.}\right)
\\
- \sum_{\substack{s,s'=\Up,\Dn \\ a=L,R \\ \tau=\Upt,\Dnt}} t_{a\tau} \left[c^\dag_{s\tau} \left(\ee^{-i a \alpha_\tau\sigma_y}\right)_{ss'}d_{a s' \tau} + {\rm H.c.} \right].
\end{multline}
In this expression we use fermionic creation and annihilation operators $d^\dag_{as\tau}, d_{as\tau}$ to represent the degrees of freedom of the external dots [indicated for simplicity by $D_{1,\ldots 4}$ in Fig. \ref{fig:tetron-schematic}(a)]. The label $a=L,R$ distinguishes the dots on the left and right side of the central superconducting island, and we respectively consider $a=+1,-1$ in the definition of the tunneling terms in the last line of Eq. \eqref{hamtot}. The label $s$ refers to the spin of the electrons. The energy level of each dot can in principle be tuned independently, however, hereafter, we consider symmetric configurations where all the dots are equivalent and described by the same chemical potential $\mu_{\rm D}$. Additionally, we include also an out-of-plane Zeeman splitting $gB_z$, which is instrumental to split the spin degeneracies in the dots and obtain poor man's Majorana modes in the non-interacting limit. We assume that the Zeeman splitting affects also the proximitized segments of the nanowires [labelled by $PD_{\tau}$ in Fig. \ref{fig:tetron-schematic} (a)], with a different Land\'e factor $g_{\rm SC}$. The electronic states in the central nanowire segments are described by the ladder operators $c^\dag_{s\tau},c_{s\tau}$ and are subject to a proximity induced s-wave pairing $\Delta$ associated with the superconducting phase operator $\ee^{i\varphi}$ that decreases the number of Cooper pairs $N_{CP}$ by one, $\left[N_{CP},\ee^{i\varphi}\right]=-\ee^{i\varphi}$. $\mu_{\rm SC}$ is an effective chemical potential for these electrons and we assume that it can be tuned by a suitable backgate (see Fig. \ref{fig:tetron-schematic}). 

The superconducting segments in each wire are tunnel coupled with the external dots with amplitudes $t_{a\tau}$, which can be controlled by cutter gates. These tunneling processes are characterized by a spin rotation determined by the Rashba spin-orbit coupling of the wires. We consider a rotation around the $y$ axis by effective angles $\pm \alpha_\tau$, depending on the direction of the tunneling. $\alpha_\tau$ is proportional to the spin-orbit momentum of nanowire $\tau$. 

Concerning the interactions $H_{\rm c}$, we focus on a scenario in which the Zeeman term $gB_z$ is strong enough to suppress the population of the spin $\Up$ states in the external dots, such that also doubly occupied states are prevented in the low-energy sector of the model. Therefore, we preliminarily disregard the effects of the onsite interactions of the four external dots and consider exclusively the charging energy of the intermediate superconducting island:
\begin{equation} \label{charging}
H_{\rm c} = E_C \left(N-n_g\right)^2\,,
\end{equation} 
where the operator $N= 2N_{CP} + \sum_{s\tau} c^\dag_{s\tau}c_{s\tau}$ describes the excess charge of the superconducting island, including the contribution given by the number of Cooper pairs $N_{CP}$ and the occupations $c^\dag_{s\tau}c_{s\tau}$ of the proximitized segments of the nanowires. The operator $N$, therefore, increases by 1 when an electron hops from one of the external dots into the central SC region. However, $N$ is left unchanged by the $\Delta$ pairing term in the Hamiltonian \eqref{hamtot}, since this term annihilates one Cooper pair to populate two electron states in the proximitized segments. The parameter $n_g= \sum_i C_{gi} V_{gi}/e$ defines the induced charge of the superconducting island and we assume that it can be tuned by suitable electrostatic side gates at voltage $V_{gi}$ (see Fig.~\ref{fig:tetron-schematic}), while the capacitances $C_{gi}$ are fixed by the geometry.

The choice in Eq. \eqref{charging} provides an approximate description of the electrostatic interactions of the poor man's tetron. A more refined modelling of the interactions can be obtained by introducing a capacitance matrix that accounts also for the crossed repulsion between electrons in the dots and central island, thus considering additional interaction terms proportional to $N d^\dag_{as\tau} d_{as\tau}$ (see the related analysis for two-dot systems in Ref. \cite{SoutoBaran}). These interactions do not yield qualitative changes in the following analysis as long as they do not introduce a strong anisotropy between the four dots. We also observe that the physics of the topological tetron is recovered in the limit of strong cross-capacitance among all the dots and between the dots and the superconducting islands. This situation is hardly achievable in hybrid semiconductor-superconductor systems where the dot charges are easily screened by the environment and the central island; however it would correspond to the limit in which the charging energy in Eq. \eqref{charging} includes not only the charge of the superconducting island, but also the charge of the four dots. In this respect the introduction of a general capacitance matrix of the device allows for the interpolation between the poor man's tetron, where electrostatic interactions are limited to the central island, to the tetron, where the charge distribution between dots and superconductor is irrelevant.

The total Hamiltonian $H$ of the poor man's tetron separately preserves the fermionic parities of the upper and lower wires:
\begin{equation} \label{parity}
P_\tau = \left(-1\right)^{\sum_{a,s} d^\dag_{as\tau} d_{as\tau} + \sum_s c^{\dag}_{s\tau} c_{s\tau}}.
\end{equation}
As a consequence, these two conserved quantities define four independent symmetry sectors.
Additionally, the total number $N_t=N+\sum_{a,s,\tau} d^\dag_{as\tau} d_{as\tau}$ of electrons in the system is conserved as well and $(-1)^{N_t}=P_\Upt P_\Dnt$. In the following we will adopt these quantum numbers to label the related sectors of the states of the system.

\section{Perturbative regime: Crossed Andreev reflection and elastic cotunneling} \label{sec:pert}

To gain a qualitative understanding of the poor man's tetron and its low-energy spectrum it is useful to analyze a perturbative regime in which, despite the presence of the charging energy, it is still possible to define elastic cotunneling processes and CARs between the external dots on the same nanowire.

\begin{figure}[t]
\centering
\includegraphics[width=1.0\columnwidth]{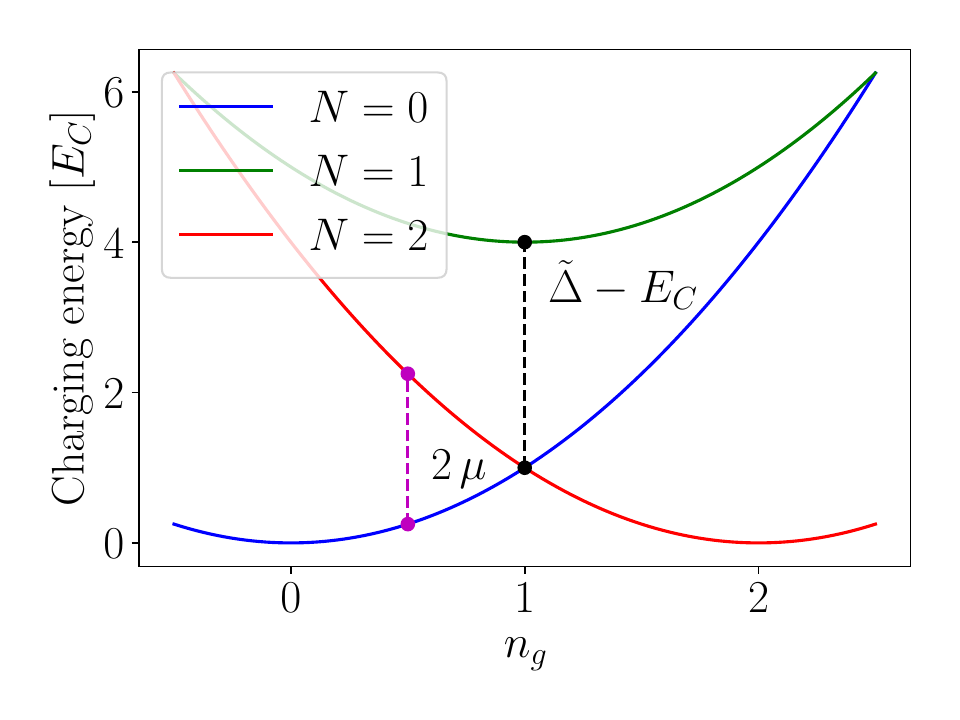}
\caption{Energy of the states of the superconducting island for $N=0,1,2$ excess electrons as a function of $n_g$. The parabolas describe the charging energy in Eq. \eqref{charging}. For odd particle numbers the energy is lifted by the SC gap $\tilde{\Delta}$. The energy difference between $N=2$ and $N=0$ states must be counterbalanced by the dot potential for optimal CAR processes (purple dashed line).}
\label{fig:Charging-parabolas}
\end{figure}

To this purpose, we extend to the poor man's tetron the analysis presented in Ref. \cite{SoutoBaran} for two dots interacting via a floating superconducting island. It is instructive to consider the energies of the states of the superconducting island, labelled by the particle number $N$, as a function of the induced charge $n_g$. For the sake of simplicity, we consider a physical setup in which the superconductor screens the magnetic field in the proximitized segments of the nanowires, such that we impose $g_{\rm SC}=0$ in the Hamiltonian \eqref{hamtot}. The general case with $g_{\rm SC}\neq 0$ is discussed in Appendix \ref{app:pert}. For $g_{\rm SC}=0$, the energy of the Bogoliubov quasiparticles in the proximitized region of the nanowires is given by:
\begin{equation}
\tilde{\Delta} = \sqrt{\Delta^2 + \mu_{\rm SC}^2}\,.
\end{equation}
We focus on the regime $E_C < \tilde{\Delta}$, where the low energy states of the superconducting island are all characterized by an even number of electrons $N$. In this situation the energy cost for occupying any Bogoliubov mode is indeed larger than the difference in charging energy between even and odd states, even for the extreme case with $n_g$ being an odd integer (see Fig.~\ref{fig:Charging-parabolas}).
This strong pairing regime allows for a simple perturbative description in terms of elastic cotunneling processes and CARs when the tunneling amplitudes $t$ are considerably smaller than $\tilde{\Delta} -  E_C$.  

The Hamiltonian \eqref{hamtot} of the poor man's tetron implies that the elastic cotunneling processes amount to the coherent tunneling of an electron from one external dot to the other in the same nanowire, mediated by a Bogoliubov state. In this process, the state of the central superconducting island is left untouched. 
The CAR, instead, corresponds to a process in which a pair of electrons coming from opposite dots in the same nanowire enter the superconducting island by forming a Cooper pair. 

In the following, we assume for simplicity that the spin-orbit coupling $\alpha$ is the same in both nanowires and all tunneling amplitudes $t_{a,\tau}$ have the same value $t$, as depicted in Fig. \ref{fig:tetron-schematic}(b). This corresponds to devices symmetric under the exchange of the two nanowires, such that cotunneling and CAR amplitudes do not depend on the $\tau$ index.

The description of cotunneling and CAR processes is further simplified in a regime of strong Zeeman splitting $gB_z > t$, which offers a double-nanowire counterpart to the hybrid poor man's Majorana devices in Ref.~\cite{SoutoBaran}. In this spin-polarized limit the low-energy behavior of the poor man's tetron can be described by the effective Hamiltonian:
\begin{multline} \label{hameff}
H_{\rm eff} = \\
-\left[t_{\rm COT}(N) \sum_{\tau} d^\dag_{R\tau} d_{L\tau} +\Delta_{\rm CAR} \ee^{-i\varphi} \sum_{\tau} d_{R\tau} d_{L\tau} + {\rm H.c.}\right] 
\\ +\left[\mu_{\rm D} - \frac{t_{\rm COT}(N)}{\cos 2\alpha}-gB_z\right] \sum_{a,\tau} d^\dag_{a\tau}d_{a\tau} + E_C \left(N-n_g\right)^2\,,
\end{multline}
where we dropped the $\Dn$ index for the spin degree of freedom. Here we introduced the elastic cotunneling amplitudes $t_{\rm COT}$, which, in general, depend on the charge $N$ of the superconducting island, and the CAR amplitude $\Delta_{\rm CAR}$.
Moreover, in Eq. \eqref{hameff} the bare potential $\mu_{\rm D}$ in Eq. \eqref{hamtot} is shifted by both the Zeeman energy and a second-order contribution $-t_{\rm COT}(N)/\cos{2\alpha}$ (see Appendix \ref{app:pert}). For the sake of simplicity, in the following we will neglect the $N$ dependence of this correction as it does not qualitatively affect our results. Therefore we simply label by 
\begin{equation}
\mu = \mu_D -gB_z - \frac{\bar{t}_{\rm COT}}{\cos{2\alpha}}
\end{equation}
an effective average energy level of the dots given by a suitable average cotunneling amplitude $\bar{t}_{\rm COT}$.

Hereafter we also restrict our analysis to induced charges $0\le n_g \le 2$. For different values of $n_g$ the physical properties repeat periodically by adjusting the number of Cooper pairs $N_{CP}$. For $0\le n_g \le 2$, the two lowest energy states of the superconducting island have charge $N=0$ and $N=2$ and correspond to the two lowest parabolas in Fig.~\ref{fig:Charging-parabolas}. CAR processes between the dots and the island allow for transitions between these states and, to significantly affect the properties of the system, they require these states to be approximately degenerate. Hence, their charging energy difference must be balanced by a suitable energy $\mu$ of the external dots. This corresponds to the condition $H_{\rm c}(N=0) + 2\mu = H_{\rm c}(N=2)$, depicted in Fig.~\ref{fig:Charging-parabolas} and yielding:
\begin{equation}\label{tuning}
\mu = 2E_C\left(1-n_g\right).
\end{equation} 
This equation defines the tuning of $\mu$ as a function of $n_g$ required to obtain optimal CAR processes in the system, even in the presence of a strong charging energy. The condition \eqref{tuning} matches an analogous constraint for the definition of poor man's Majorana modes in two-dot devices with an intermediate floating superconducting island \cite{SoutoBaran}.
The introduction of additional cross-capacitance terms in the Hamiltonian would simply result in a redefinition of $H_c$ and Eq. \eqref{tuning}, as long as all the dots are equivalent to each other.

The general derivation of the cotunneling and CAR amplitudes is presented in Appendix \ref{app:pert}. For spin-polarized dots tuned based on the condition \eqref{tuning}, we obtain:
\begin{multline} \label{cot}
t_{\rm COT}\left(N= 1 \pm 1 \right)= \\
+t^2\cos(2\alpha) \frac{\mu_{\rm SC}\tilde{\Delta} + \mu_{\rm SC}E_C \mp 2E_C\tilde{\Delta} }{\tilde{\Delta} \left(\tilde{\Delta} + 3E_C\right)\left(\tilde{\Delta} - E_C\right)}\,,
\end{multline}
and
\begin{equation} \label{CAR2}
\Delta_{\rm CAR} = \frac{t^2 \sin\left(2\alpha\right) \sqrt{\tilde{\Delta}^2-\mu_{\rm SC}^2}}{\tilde{\Delta}\left(\tilde{\Delta}-E_C\right)}\,.
\end{equation}
We observe, in particular, that the dependence of the cotunneling amplitude on the occupation of the superconducting island is approximately proportional to $t^2 E_C \cos(2\alpha)/\tilde{\Delta}^2$, thus vanishing for small $E_C/\tilde{\Delta}$. The CAR amplitude, instead, is unaffected by $N$.

In the following sections we will primarily consider setups in which the tuning condition \eqref{tuning} is fulfilled. We stress, however that deviations from this optimal chemical potential that are small compared to the CAR and cotunneling amplitudes do not yield major effects on the following analysis.

\section{Low-energy states in the weak-tunneling regime} \label{PMtetron}

 \begin{table*}[t]
\begin{tabular}{|c|c|}
\hline
\textbf{Assumption} & \textbf{Effect} \\
\hline
$\tilde{\Delta} - g_{\rm SC} B_z > E_C$ & CAR processes are well defined at low energy \\
\hline
$\tilde{\Delta} - g_{\rm SC} B_z - E_C \gg t \gg |t_{\rm COT}|, |\Delta_{\rm CAR}|$ & Perturbative tunneling regime \\
\hline
$gB_z \gg |t_{\rm COT}|, |\Delta_{\rm CAR}|$ & Polarization of the dots \\
\hline
$E_C \gtrsim |t_{\rm COT}|, |\Delta_{\rm CAR}|$ & The lowest energy states display $N_{t} \in [0,6]$\\
\hline
$\mu \sim 2E_C\left(1-n_g\right)$ & Degeneracy of the $N=0$ and $N=2$ charge states\\
\hline
\end{tabular}
\caption{Assumptions adopted for the perturbative analysis of the poor man's tetron}
\label{table}
\end{table*}

In the absence of coupling to external leads, the total number of electrons in the poor man's tetron is preserved. For $n_g \in [0,2]$ and a sizeable $E_C$ (comparable with CAR and cotunneling amplitudes), the states with the lowest energy display a total electron number $N_{t}=N+\sum_{a,\tau} d^\dag_{a\tau} d_{a\tau} \in \left[0,6\right]$. Therefore, in the following, we will focus on states with these total particle numbers. 

Given the total number $N_t$ of electrons, and thereby the total parity of the system $P_\Upt P_\Dnt$, there are eight possible states that can be distinguished by the occupation numbers $n_{a\tau} = d^\dag_{a\tau}d_{a\tau}=0,1$. These eight states are, in turn, subdivided into two sets of four states with $P_\Dnt=\pm 1$.

Consequently, the Hamiltonian $H_{\rm eff}$ can be decomposed into $4 \times 4$ subblocks, with two subblocks for each total electron number $N_t$.  Their diagonalization allows us to determine the structure of the low-energy states of the poor man's tetron.

The $4 \times 4$ Hamiltonian blocks acquire, in general, four possible forms determined by the parities $P_\Upt$ and $P_\Dnt$ which establish whether the dynamics of the related sector in each wire is dictated by CAR or cotunneling processes. Therefore, in general, it will be convenient to label the sectors with the notation $(N_t, P_\Dnt)$, where the total number of particles $N_t$ determines the total parity of the system $P_\Upt P_\Dnt$, and $P_\Dnt$ further specifies which of the four $(P_\Upt,P_\Dnt)$ sector is addressed.

In the following, we analyze the ground state of each parity block to understand the low-energy behavior of the system. Table \ref{table} summarizes the main assumptions for the validity of the following perturbative results and their effect; Appendix \ref{app:sectors} presents the effective Hamiltonians in each sector and the related ground state energies.

\subsection{The odd parity states}

The odd parity sectors $(N_t=2m+1,P_\Dnt=+)$ and $(N_t=2m+1,P_\Dnt=-)$ are affected by both CAR and cotunneling processes. When the low-energy Hamiltonian \eqref{hameff} is invariant under the exchange of the two nanowires, $\Upt \leftrightarrow \Dnt$, these two subspaces are equivalent. This emerging symmetry does not require a perfect physical equivalence of the two nanowires; it corresponds instead to a tuning of the poor man's tetron such that the differences of the CAR and cotunneling amplitudes between the two wires are negligible. Even if the parameters in the microscopic Hamiltonian $H_{\rm sp}$ (for instance $\Delta$ or $g_{\rm SC}$) are different for the two nanowires in realistic devices, the control of the surrounding electrostatic gates enables several strategies to compensate these discrepancies at the level of $H_{\rm eff}$ (see Appendix \ref{app:pert}). 

If we impose the tuning condition \eqref{tuning}, the odd sectors with $N_t=3$ are the only subspaces in which all the CAR processes connect degenerate charge states (see Appendix \ref{app:sectors}). This implies that the eigenstates in $(N_t=3,P_\Dnt)$ display only a weak dependence on the charging energy, inherited by the CAR and cotunneling amplitudes. This scenario is close to the non-interacting limit described by poor man's Majorana modes and, for this reason, we focus our analysis on setups that favor the onset of global ground states in these sectors.

We address first the $(3,+)$ sector.
By ordering the dots occupation numbers $n_i=d^\dag_id_i$ with $i=1,\ldots,4$ as indicated in Fig.~\ref{fig:tetron-schematic}, we can express the four states in this subspace in the basis $\ket{n_1 n_2 n_3 n_4,N}$. They result: $\ket{0100,2}$, $\ket{1000,2}$, $\ket{0111,0}$ and $\ket{1011,0}$. 

We consider a region of parameter space where
$t_{\rm COT}(2) < t_{\rm COT}(0) < 0$ and $\Delta_{\rm CAR} >0$; this corresponds to small spin-orbit coupling ($0<\alpha<\pi/4$) and negative $\mu_{SC}<-2E_C\tilde{\Delta}/(\tilde{\Delta} + E_C)$. In this case the ground state assumes the form:
\begin{multline} \label{o1}
\ket{3+} = \frac{1}{\sqrt{2}}\left[\cos\theta \left( \ket{0100,2} - \ket{1000,2}\right) \right.\\
\left.+ \sin\theta\left( \ket{0111,0} -\ket{1011,0}\right) \right].
\end{multline}
Given the symmetry between the two nanowires, the sector $(3,-)$ displays a ground state with the same energy and the form:
\begin{multline} \label{o2}
\ket{3-} = \frac{1}{\sqrt{2}}\left[\cos\theta \left( \ket{0001,2} - \ket{0010,2}\right) \right.\\
\left.+ \sin\theta \left( \ket{1101,0} -\ket{1110,0}\right) \right].
\end{multline}
The angle $\theta$ is a parameter that can be experimentally controlled through the energy level $\mu_D$ of the dots. When imposing the degeneracy condition \eqref{tuning}, it reads:
\begin{equation} \label{theta}
\theta= \frac{1}{2}\arctan \frac{2\Delta_{\rm CAR}}{t_{\rm COT}(0) - t_{\rm COT}(2)}\,.
\end{equation}
In the general case, instead, the dot potential $\mu_D$ determines the denominator in Eq. \eqref{theta}.

In the limit $E_C \to 0$, the two cotunneling amplitudes $t_{\rm COT}(0)$ and $t_{\rm COT}(2)$ become equal [see Eq. \eqref{cot}]. Therefore, based on Eq. \eqref{theta}, $\theta$ tends to $\pi/4$ and the odd states $\ket{3,\pm}$ become a superposition with equal amplitudes of their four components, hence recovering the non-interacting result expected for two two-site Kitaev chains with vanishing onsite potential.

The odd parity sectors with $N_t \neq 3$ present ground states with the same form but their states display different charging energies (see Appendix \ref{app:sectors}) which yield an explicit dependence of the parameter $\theta$ on $E_C$.

We finally mention that there is a special line in parameter space, obtained by imposing $\mu_{SC}=0$ and fulfilling the degeneracy condition \eqref{tuning}, such that $t_{\rm COT}(2)=-t_{\rm COT}(0)$ and the ground state of each sector $(3,P_\Dnt)$ becomes two-fold degenerate. We emphasize, however, that our perturbative results indicate that the global ground state of the tetron does not belong to the $N_t=3$ sectors in the limit $\mu_{SC}\to 0$.

\subsection{The even parity states}

The low-energy sectors with even fermionic parity that compete with the odd states to be the global ground states of the poor man's tetron are characterized by $N_t=2$ or $N_t=4$ total electrons for $0\le n_g\le 2$.

The even parity sectors $(N_t=2m,P_\Dnt=+)$ and $(N_t=2m,P_\Dnt=-)$ are inequivalent since the former has an effective Hamiltonian that depends on CAR processes only, whereas the energies of the latter depend only on elastic cotunneling processes. Their Hamiltonians are presented in Appendix \ref{app:sectors}.

The sector $(2m,+)$ is the only sector involving states with three different values of the charge in the superconducting island. For $N_t=2,4$, the tuning of the potential of the external dots allows us to set two of these three charge states to be degenerate. The third acquires a higher energy. As a result, the ground state of $(2m,+)$ displays, in general, a form that considerably detaches from the non-interacting limit $E_C\to 0$.

In the strongly interacting case $E_C \gg \Delta_{\rm CAR}$, the related ground states result: 
\begin{multline}
\ket{4+} \xrightarrow{E_C \gg \Delta_{\rm CAR}} \frac{1}{2} \ket{0011,2} 
\\ + \frac{1}{2} \ket{1100,2} + \frac{1}{\sqrt{2}} \ket{1111,0} \label{even1_largeEc}\,,
\end{multline}
\begin{multline}
\ket{2+} \xrightarrow{E_C \gg \Delta_{\rm CAR}} \frac{1}{2} \ket{0011,0} 
\\ + \frac{1}{2} \ket{1100,0} + \frac{1}{\sqrt{2}} \ket{0000,2} \label{even1_largeEc2}\,.
\end{multline}
The sector $(2m,-)$, instead, has eigenstates determined exclusively by the cotunneling processes and all its states display a charge $N_t-2$ in the superconducting island. For these reasons, its ground state does not depend on $E_C$ and reads:
\begin{multline} \label{e2}
\ket{2m-} = \frac{1}{2}\left[\ket{0101,N_t-2} - \ket{0110,N_t-2} \right.\\
\left.- \ket{1001,N_t-2}+ \ket{1010,N_t-2}\right]\,.
\end{multline}
The choice of the signs in Eq. \eqref{e2} corresponds to the case $t_{\rm COT}(N_t-2)<0$. The case $t_{\rm COT}(N_t-2)>0$ does not present any qualitative difference and the general scenario is considered in Appendices \ref{app:sectors} and \ref{app:exci}.

\subsection{Effects of the charging energy}

The fermionic parities $P_\tau$ are symmetries of the Hamiltonian of the poor man's tetron irrespective of whether the superconducting island is floating or grounded. Thus, the previous sectors have a clear counterpart also in the non-interacting limit $E_C,\mu \to 0$, even when the total electron number is not conserved.

Therefore also in the non-interacting case we can distinguish four low-energy ground states which, for $t_{\rm COT}<0$ and $\Delta_{\rm CAR}>0$, have wavefunctions:
\begin{align*}
\ket{\text{e}+} =& \frac{1}{2}\left(\ket{00}+\ket{11}\right)_{\Upt} \otimes \left(\ket{00}+\ket{11}\right)_{\Dnt}\,,\\
\ket{\text{e}-} =& \frac{1}{2}\left(\ket{01}-\ket{10}\right)_{\Upt} \otimes \left(\ket{01}-\ket{10}\right)_{\Dnt}\,,\\
\ket{\text{o}+} =&  \frac{1}{2}\left(\ket{01}-\ket{10}\right)_{\Upt} \otimes \left(\ket{00}+\ket{11}\right)_{\Dnt}\,, \\
\ket{\text{o}-} =&  \frac{1}{2} \left(\ket{00}+\ket{11}\right)_{\Upt}\otimes \left(\ket{01}-\ket{10}\right)_{\Dnt}\,.
\end{align*}
Their energies are:
\begin{align*}
E^{(\text{e}+)} =& -\left|\Delta_{\rm CAR,\Upt}\right|-\left|\Delta_{\rm CAR,\Dnt}\right|\,,\\
E^{(\text{e}-)} =& -\left|t_{\rm COT,\Upt}\right|-\left|t_{\rm COT,\Dnt}\right|\,,\\
E^{(\text{o}+)} =&  -\left|t_{\rm COT,\Upt}\right|-\left|\Delta_{\rm CAR,\Dnt}\right|\,, \\
E^{(\text{o}-)} =& -\left|\Delta_{\rm CAR,\Upt}\right|-\left|t_{\rm COT,\Dnt}\right|\,.
\end{align*}
For systems symmetric under $\Upt \leftrightarrow \Dnt$, the odd ground states are degenerate and their energy, $-|t_{\rm COT}| -|\Delta_{\rm CAR}|$, lies in between the energies of the even ground states. 

Additionally, a four-fold degeneracy of the system is obtained at the sweet spot corresponding to $\left|\Delta_{\rm CAR,\tau}\right| = \left|t_{\rm COT,\tau}\right|$ for both nanowires. This corresponds to the formation of four poor man's Majorana modes \cite{Liu2022}.

When the particle number is conserved, the four sectors are further split based on the total number of electrons $N_t$. Subsectors of the same kind, however, remain degenerate for $E_C \to 0$.

In floating devices with $E_C>0$, only the state $\ket{\text{e}-}$ maintains its non-interacting form. 
When imposing the degeneracy of states with charge $N=0$ and $N=2$ in the superconducting island via the tuning \eqref{tuning}, the odd states with $N_t=3$ acquire a correction in the coefficients of their wavefunctions that is represented by the displacement of $\theta$ from $\pi/4$ in Eqs. \eqref{o1} and \eqref{o2}. This correction, at leading order, is linear in $\left[t_{\rm COT}(0) - t_{\rm COT}(2)\right]/\Delta_{\rm CAR}$, thus in $E_C/\Delta$ [see Eqs. (\ref{cot},\ref{CAR2})].

\begin{figure}[t]
\centering
\includegraphics[width=0.9\columnwidth]{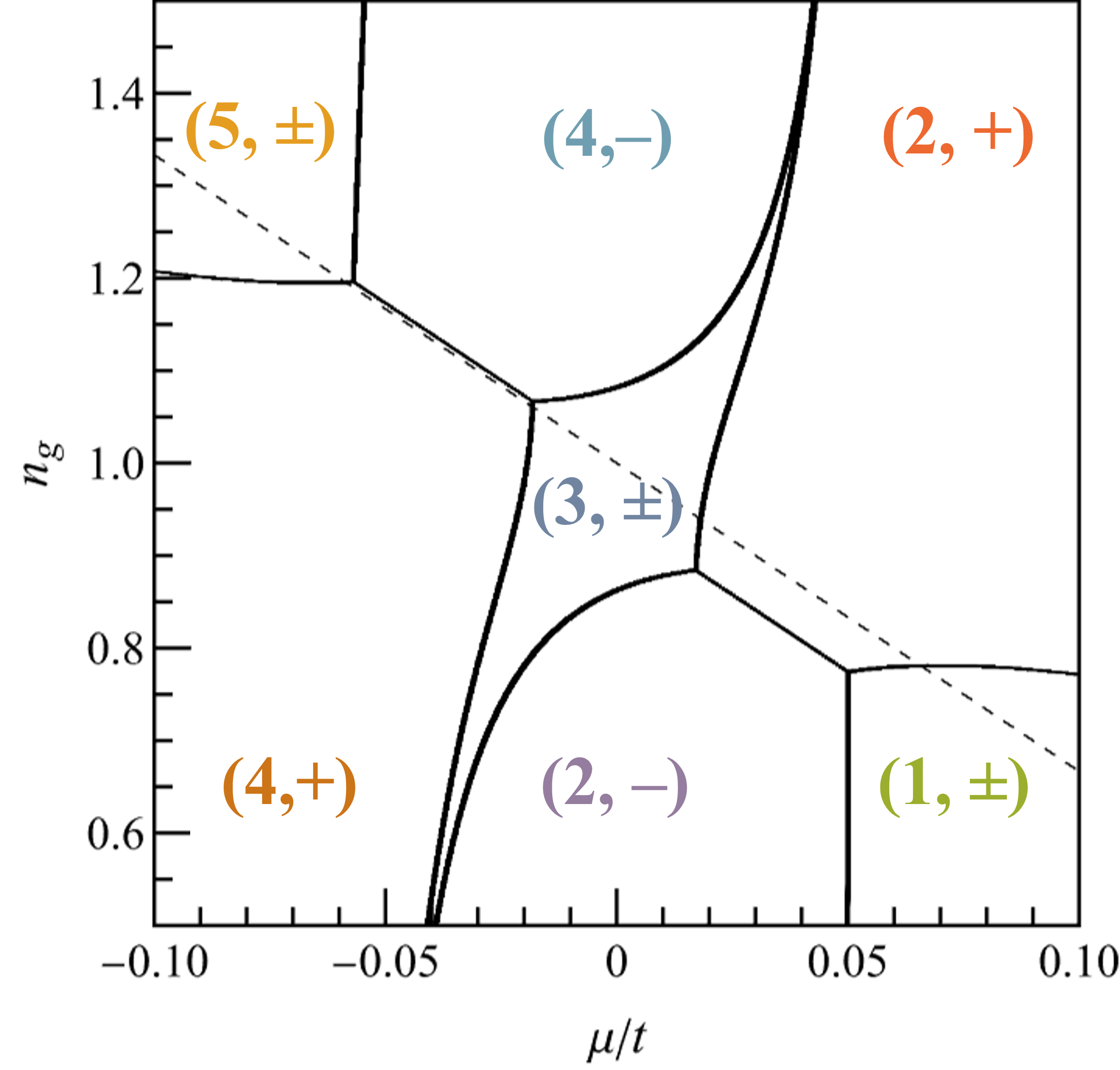}
\caption{Alternation of the parity sectors that express the global ground state of the poor man's tetron as a function of $n_g$ and $\mu$. The parameters are: $\Delta=4t$, $\mu_{SC}=-5t$, $E_C=0.15t$, $\alpha=\pi/6$. The dashed line corresponds to the constraint \eqref{tuning} and the energies in Fig.~\ref{fig:spectrum}. The labels indicate the particle number $N_t$ and the parity $P_\Dnt$.}
\label{fig:sectors}
\end{figure}

The states in odd sectors with $N_t \neq 3$, as well as the states in the $(2m,+)$ sectors, instead, are more affected by $E_C$ given the different energy costs of their charge configurations. Concerning the states $\ket{4+}$ and $\ket{2+}$, this can be easily seen by comparing Eqs. (\ref{even1_largeEc},\ref{even1_largeEc2}), which display only three components, with the non-interacting form $\ket{\text{e}+}$, which, instead, is the equal superposition of four terms (see App. \ref{app:sectors} for further details). For this reason, the four poor man's Majorana modes are not stable, in general, under the introduction of the charging energy.

Despite this, however, it is still possible to identify regions in the parameter space of the poor man's tetron where this device displays degenerate ground states and may approach the physics of the topological tetron. To this purpose, we need to identify regimes of the parameters in which the ground states of the odd sectors with $N_t=3$ constitute the lowest energy states of the poor man's tetron and the emerging nanowire symmetry $\Upt \leftrightarrow \Dnt$ is achieved. Importantly, the appearance of global ground states with an odd fermionic parity requires a non-vanishing charging energy because, in the non-interacting limit, the energy of the odd ground states lies in between the energies of the even states.

The even sectors with $N_t=2,4$ are the main competitors to establish the global ground state for $n_g\sim 1$. This can be seen in the example reported in Fig.~\ref{fig:sectors} where we illustrate which sector expresses the global ground state as a function of $n_g$ and $\mu$ for a choice of parameters such that at $\mu=0$ and $n_g=1$ there is a degenerate odd ground state belonging to the sectors $\left(3,\pm\right)$. Ground states with even fermionic parity surround the central region with the odd ground states at $N_t=3$.

\begin{figure}[t]
\includegraphics[width=0.9\columnwidth]{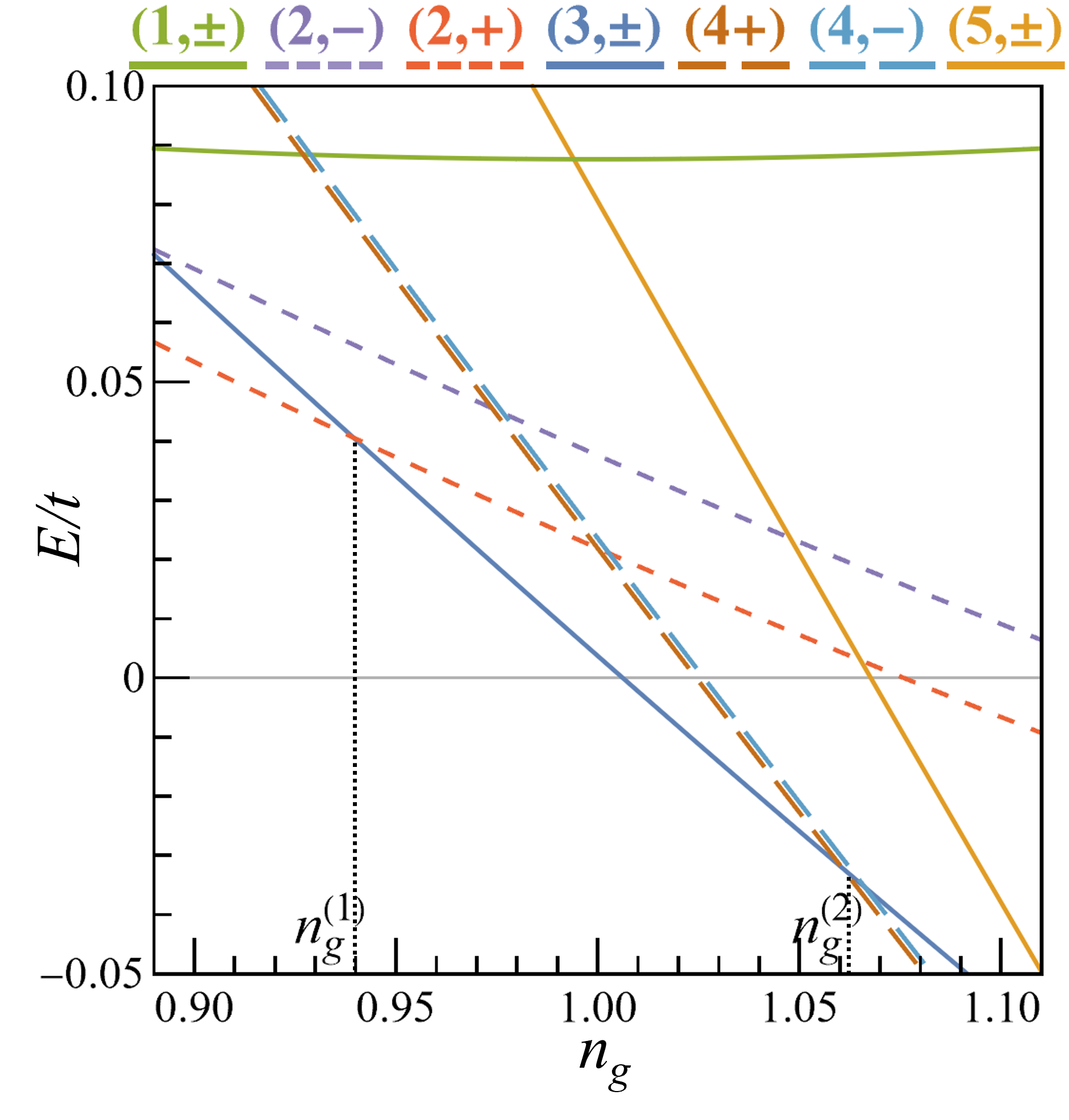}
\caption{Energies of the ground states of the parity sectors  of the poor man's tetron with $N_t=1,\ldots,5$ in the perturbative regime as a function of the induced charge $n_g$. The labels indicate the particle number $N_t$ and the parity $P_\Dnt$. The chemical potential $\mu(n_g)$ is chosen to fulfill Eq. \eqref{tuning}. The other parameters are $\Delta=4t$, $\mu_{SC}=-5t$, $E_C=0.15t$, $\alpha=\pi/6$. For this choice of parameters $\Delta_{\rm CAR}\approx 0.087t$, $t_{\rm COT}(0) \approx -0.056t$ and $t_{\rm COT}(2) \approx -0.063t$.}
\label{fig:spectrum}
\end{figure}

To gain a qualitative intuition of the different behavior of the ground state energies of even and odd sectors let us first observe that, due to the different electron number, the even states with $N_t=2,4$ are affected in a different way by the dot energy level and, in general, they gain a different energy contribution $\pm \mu$ with respect to the odd $N_t=3$ ground states. Fig.~\ref{fig:spectrum} displays the low-energy spectrum of the poor man's tetron close to $n_g=1$ when the potential of the external dots fulfills the constraint \eqref{tuning}. The chosen parameters correspond to the dashed cut in Fig.~\ref{fig:sectors} and the ground state displays $N_t=3$ electrons at $n_g=1$ (blue line in Fig.~\ref{fig:spectrum}). Eq. \eqref{tuning} implies that the $\pm \mu$ energy shift of the even states is translated into a different slope of the even ground states energies with respect to $n_g$ (compare the blue line with the dashed and dotted lines). Furthermore, the sign of $\mu$ changes at $n_g=1$. Therefore even states with $N_t=2$ (dashed lines in Fig.~\ref{fig:spectrum}) are favored for lower values of $n_g$ and $\mu>0$, whereas even states with $N_t=4$ (dotted lines) are favored for higher values of $n_g$ and $\mu<0$. 

To better understand the spectrum close to $n_g=1$, let us consider the regime with $\tilde{\Delta} \gg E_C \gg |\Delta_{\rm CAR}|$. In this case, the energy differences between the ground states of the sectors with $N_t=2,4$ (at energies $E^{(N_t=2,4,\pm)}$) and the odd sectors $N_t=3$ (at energies $E^{(3)}$) can be roughly approximated as:
\begin{align}
&\delta E_+^{(N_t=3\pm1)} \equiv E^{(N_t=3\pm1,+)} - E^{(3)} \approx \nonumber \\  
&\qquad \pm \mu - \left(\sqrt{2}-1\right)\Delta_{\rm CAR} +\bar{t}_{\rm COT}\,, \label{dE1}\\
&\delta E_-^{(N_t=3\pm1)} \equiv E^{(3\pm1,-)} - E^{(3)} \approx \nonumber
\\ & \qquad \pm\mu - \bar{t}_{\rm COT}+ \Delta_{\rm CAR}\,, \label{dE3}
\end{align}
where the $\pm$ subscripts indicate the parity $P_{\Dnt}$ of the related even sectors. For simplicity, in the previous equations we replaced all the cotunneling amplitudes with the average value:
\begin{equation}
\bar{t}_{\rm COT}\equiv\frac{\left|t_{\rm COT}(2)+t_{\rm COT}(0)\right|}{2} \,.
\end{equation}

By setting the dot energies to the value $\mu(n_g)$ in Eq. \eqref{tuning}, the energy gaps in Eqs. \eqref{dE1} and \eqref{dE3} determine whether there exists a range of induced charges $n_g$ such that the degenerate odd states $\ket{3+}$ and $\ket{3-}$ become the ground states of the poor man's tetron.
The perturbative analysis suggests that these gaps are positive around $n_g=1$ for intermediate values of the charging energies. On one side, $E_C$ must be sufficiently larger than $\Delta_{\rm CAR}$ such that the modulus of the coefficient of $\Delta_{\rm CAR}$ in Eq. \eqref{dE1} is sufficiently smaller than 1. Eq. \eqref{dE1} corresponds indeed to the large $E_C$ limit in Eqs. (\ref{even1_largeEc},\ref{even1_largeEc2}) such that this coefficient acquires its minimum value $\sqrt{2}-1$. On the other, $E_C$ must be sufficiently small to avoid a large difference between the cotunneling amplitudes $t_{\rm COT}(0)$ and $t_{\rm COT}(2)$. Indeed, in Eqs. \eqref{dE1} and \eqref{dE3} we neglected this difference which, however, linearly displaces the two ranges of the parameter $\mu$ which respectively fulfill $\delta E_+^{(N_t=2,4)}>0$ and $\delta E_-^{(N_t=2,4)}>0$.
Therefore, to satisfy all conditions $\delta E_{\pm}^{(N_t=2,4)}>0$ we require:
\begin{equation}
|t_{\rm COT}(0)| - |t_{\rm COT}(2)| \ll \Delta_{\rm CAR} - \bar{t}_{\rm COT}\,.
\end{equation}
Under these assumptions, the onset of ground states with odd fermionic parity is favoured at $\Delta_{\rm CAR} \lesssim \sqrt{2}\bar{t}_{\rm COT}$. 
We emphasize that these conditions are derived based on perturbation theory. In Sec. \ref{sec:surrogate} we will consider the system beyond the perturbative regime and show that the poor man's tetron can display degenerate odd ground states even if these conditions are violated when considering systems with stronger tunnelling amplitudes $t$.

Concerning the choice of parameters in Fig.~\ref{fig:spectrum}, the CAR amplitude is approximately $\Delta_{\rm CAR} \approx 1.45 \bar{t}_{\rm COT}$ and the degenerate states $\ket{3+}$ and $\ket{3-}$ are ground states of the system in a range of induced charges $n_g^{(1)} < n_g < n_g^{(2)}$. This is the regime we will consider in the next section to investigate the Kondo effects emerging in the poor man's tetron.

At the points $n_g=n_g^{(1,2)}$ the ground state becomes threefold degenerate with one of the even states crossing the odd ground states. At $n_g=n_g^{(1)}$, in particular $E^{(2,+)} = E^{(3)}$: this condition corresponds to a situation in which $\mu$ counterbalances the difference between CAR and cotunneling amplitudes and matches the sweet spot discussed in Ref.~\cite{SoutoBaran} for a single-nanowire device with a floating superconducting island. For this specific value of the induced charge, the state $\ket{2-}$ is an excited state lifted in energy from the three ground states by approximately $\sqrt{2} \Delta_{\rm CAR} - 2 \left|t_{\rm COT}(0)\right|$ in the limit $E_C \gg \Delta_{\rm CAR}$.

The structure of the eigenstates close to $n_g^{(1)}$ in the strong $E_C$ limit is reminiscent of the three-fold degenerate spinful systems studied in Ref.~\cite{Bozkurt2024}, where the nanowire degree of freedom of the spin-polarized poor man's tetron plays a role similar to spin in those two-dot setups. This regime with three low-lying ground states well separated in energy from $\ket{2-}$ is the situation we address in Sec. \ref{sec:anderson} by mapping the poor man's tetron onto an Anderson impurity model.

In experimental setups, two-terminal conductance measurements performed with external leads connected to the dots on the same wire would allow for the determination of these spectral features: the gaps $\delta E_\pm$ can be measured by finite-bias spectroscopy and the charges $n_g^{(i)}$ can be determined by observing the zero-bias peak of the system. Finally, we emphasize that the observations presented so far refer only to the ground states of each of the four sectors. Concerning the excited states within each sector, however, the related gaps are approximately given by $2\Delta_{\rm CAR}$ or $2t_{\rm COT}(N)$ and are larger than the gaps $\delta E_\pm^{(N_t=2,4)}$ in our regimes of interest.

\section{The three-lead Kondo model} \label{Kondo}

The most straightforward experimental investigations that can be performed in a hybrid quantum dot / superconductor system like the poor man's tetron are based on its transport properties. Therefore, in the following, we extend our analysis of this device to include external leads. In particular, we focus on a multiterminal scenario in which the zero-temperature limit of the conductance matrix is intimately related to the onset of Kondo-like correlations between the poor man's tetron and these external electrodes, such that transport can reveal important signatures about the spectrum of the system and its low-energy states.

In the case of topologically protected tetrons, their coupling to external leads results in the topological Kondo effect \cite{Beri2012,Beri2013,Altland2013} as the tetron's Majorana modes concur to define a non-local spin 1/2 impurity. In the following we investigate to what extent the peculiar forms and correlations of the ground states discussed in the previous section may allow us to retrieve the topological Kondo effect through a tuning of the poor man's tetron, even without topological Majorana modes.

To this purpose, we extend our analysis to the following Hamiltonian:
\begin{equation} \label{dotcoupl}
H = H_{\rm sys} + H_{\rm leads} -\sum_\alpha J_\alpha l^\dag_\alpha d_\alpha + \rm{H.c.}  \,.
\end{equation}
The index $\alpha$ labels the leads [Fig.~\ref{fig:tetron-schematic}(a)], each of which is tunnel-coupled with one of the dots with amplitude $J_\alpha$. As in the previous section, we assume here that the Zeeman term in the dots is strong, such that only spin down electrons are involved in the low-energy physics of this model and, therefore, also the leads can be approximated as polarized systems. Hence, the operator $l^\dag_\alpha$ constitutes the creation operators of an electron at the extremity of lead $\alpha$. In Sec. \ref{sec:surrogate} we will present results on the more general spinful scenario.

$H$ can be interpreted as the Hamiltonian of an interacting impurity, corresponding to the poor man's tetron, coupled with external leads as in an Anderson problem. Specifically, we are interested in the regime in which the two odd ground states of the tetron [Eqs. \eqref{o1} and \eqref{o2}] constitute two degenerate states of an impurity which, as a consequence, can be treated as an effective spin 1/2 system. In the following we will consider indeed setups in which the two nanowires are equivalent at low energies, such that the odd ground states $\ket{3+}$ and $\ket{3-}$ are degenerate [see Appendix \ref{app:sectors}]. A violation of this condition would correspond to the introduction of an effective Zeeman term in the analogous Anderson model.

We can obtain insight of the Kondo problem emerging from the coupling of this effective spin 1/2 degree of freedom with the leads by considering a perturbative regime in which the parameters $J_\alpha$ are the smallest energy scale in the system, such that $|J_\alpha| \ll \delta E_\pm^{(N_t)}$ for all even sectors $(2m,\pm)$. In particular a Kondo model emerges at second order in perturbation theory and can be described by the following Hamiltonian:
\begin{equation} \label{hkondo}
H_{\rm Kondo} = \sum_{\alpha,\beta} J_\alpha  J_\beta^* \left(l^\dag_\alpha l_\beta \Upsilon^{\alpha\beta} + l_\beta l^\dag_\alpha \Xi^{\alpha\beta}\right) \,.
\end{equation}
Here $\alpha$ and $\beta$ run over the external leads and the operators $\Upsilon^{\alpha\beta}$ and $\Xi^{\alpha\beta}$ are $2\times 2$ matrices acting on the effective spin 1/2 degree of freedom defined by the ground states $\{\ket{3+},\ket{3-}\}$. By considering the low-energy subspace described in the previous section, these matrices read:
\begin{multline} \label{Upsilon}
\Upsilon^{\alpha \beta} = \sum_{\substack{n\\ a,b,c=\pm}}  \frac{-1}{E^{(4,c)}_n - E^{(3)}} \\
\ket{3a} \bra{3a} d_\alpha \ket{4,c,n} \bra{4,c,n} d^\dag_\beta \ket{3b} \bra{3b}\,,
\end{multline}
\begin{multline} \label{Xi}
\Xi^{\alpha \beta}=\sum_{\substack{n\\ a,b,c=\pm}}\frac{-1}{E^{(2,c)}_n - E^{(3)}} \\ \ket{3a} \bra{3a} d^\dag_\beta \ket{2,c,n}  \bra{2,c,n} d_\alpha \ket{3b} \bra{3b}\,.
\end{multline}

In this expressions the index $n$ runs over the four eigenstates in each even sector of the Hamiltonian $H_{\rm eff}$ in Eq. \eqref{hameff}. The energy $E_n^{(N_t,c)}$ corresponds to the eigenenergy of the $n^{\rm th}$ eigenstate of the even sector $(N_t,c)$, labelled by $\ket{N_t,c,n}$ (see Appendix \ref{app:sectors}). $E^{(3)}$ is the energy of the odd ground states with $N_t=3$ electrons [Eq. \eqref{Eodd}]. The sums in the definition of $\Upsilon$ and $\Xi$ run over the even states with $N_t=4$ and $N_t=2$ electrons respectively. 

For the sake of simplicity, in the following subsections we will consider only the ground states $n=1$ of the four involved even sectors. States with higher $n$ have indeed higher energies and do not qualitatively affect the formulation of the Kondo problem emerging from the Hamiltonian \eqref{hkondo}. The quantitative effect of the excited states with $n>1$ is outlined in Appendix \ref{app:exci} and it amounts to a variation of the order of $\sim 4\%$ in the definition of the $\Upsilon$ and $\Sigma$ matrices for the parameters considered in Fig.~\ref{fig:spectrum}. 

When restricting to the ground states ($n=1$), and considering the limit $E_C \gg \Delta_{\rm{CAR}}$ in Eq. (\ref{even1_largeEc}), the matrix elements of the operator $d^\dag_\beta$ in Eq. \eqref{Upsilon} read:
\begin{multline} \label{ddagm}
\sum_{b, c = \pm} \bra{4,c,1} d^\dag_\beta \ket{3b} \, \ket{4,c,1}\bra{3b} = \\
\frac{\cos\theta}{\sqrt{2}} \begin{pmatrix}
\xi_4 (\delta_{\beta, 1} + \delta_{\beta, 2}) & \xi_4 (\delta_{\beta, 3} + \delta_{\beta, 4}) \\
 \delta_{\beta, 3} - \delta_{\beta, 4} & -\delta_{\beta, 1}  + \delta_{\beta, 2} \\
\end{pmatrix}\,.
\end{multline}
Here we introduced the numerical parameter $\xi_4$ for the sector $N_t=4$:
\begin{equation}
\xi_4 = \frac{1 + \sqrt{2} \tan \theta}{2}\,.
\end{equation}
This parameter codifies the ratio of the transitions from the odd ground states to the even ground states of the sectors $(4,c)$. The angle $\theta$ is defined in Eq. \eqref{theta}. We observe that $\xi$ is related to the deviations of the four many-body ground states from their non-interacting limits: $E_C$ dictates indeed the angle $\theta$ and the factor $\sqrt{2}$ is obtained in the strong interaction limit [Eq. \eqref{even1_largeEc}].
In the non-interacting case ($E_C=0$), $\xi_4=1$.

In an analogous way the matrix elements of $d_\alpha$ in Eq. \eqref{Xi} read:
\begin{multline} \label{dm}
\sum_{b, c = \pm} \bra{2,c,1} d_\alpha \ket{3b} \, \ket{2,c,1}\bra{3b} = \\
\frac{\sin\theta}{\sqrt{2}} \begin{pmatrix}
\xi_2 (\delta_{\alpha, 2} - \delta_{\alpha, 1}) & \xi_2 (\delta_{\alpha, 4} - \delta_{\alpha, 3}) \\
 -\delta_{\alpha, 3} - \delta_{\alpha, 4} & \delta_{\alpha, 1}  + \delta_{\alpha, 2} \\
\end{pmatrix}\,,
\end{multline}
with
\begin{equation}
\xi_2 = \frac{1 + \sqrt{2} \cot \theta}{2}\,.
\end{equation}
Based on the previous equations we obtain:
\begin{widetext}
\begin{equation}
\Upsilon^{\alpha \beta}= -\frac{\cos^2\theta}{2}  
 \begin{pmatrix}
\frac{\xi^2_4 (\delta_{\alpha, 1} + \delta_{\alpha, 2}) (\delta_{\beta, 1} + \delta_{\beta, 2})}{\delta E_+^{(4)}}  + \frac{(\delta_{\alpha, 3} - \delta_{\alpha, 4}) (\delta_{\beta, 3} - \delta_{\beta, 4})}{\delta E_-^{(4)}}  &
\frac{\xi^2_4(\delta_{\alpha, 3} + \delta_{\alpha, 4}) (\delta_{\beta, 1} + \delta_{\beta, 2})}{\delta E_+^{(4)}}  - \frac{(\delta_{\alpha, 1} - \delta_{\alpha, 2}) (\delta_{\beta, 3} - \delta_{\beta, 4})}{\delta E_-^{(4)}} \\
\frac{\xi^2_4(\delta_{\alpha, 1} + \delta_{\alpha, 2}) (\delta_{\beta, 3} + \delta_{\beta, 4})}{\delta E_+^{(4)}}  - \frac{(\delta_{\alpha, 3} - \delta_{\alpha, 4}) (\delta_{\beta, 1} - \delta_{\beta, 2})}{\delta E_-^{(4)}} &
\frac{\xi^2_4(\delta_{\alpha, 3} + \delta_{\alpha, 4}) (\delta_{\beta, 3} + \delta_{\beta, 4})}{\delta E_+^{(4)}}  + \frac{(\delta_{\alpha, 1} - \delta_{\alpha, 2}) (\delta_{\beta, 1} - \delta_{\beta, 2})}{\delta E_-^{(4)}} \\
\end{pmatrix},
\end{equation}
\begin{equation}
\Xi^{\alpha \beta}= -\frac{\sin^2\theta}{2 }  
 \begin{pmatrix}
\frac{\xi^2_2(\delta_{\alpha, 1} - \delta_{\alpha, 2}) (\delta_{\beta, 1} - \delta_{\beta, 2})}{\delta E_+^{(2)}}  + \frac{(\delta_{\alpha, 3} + \delta_{\alpha, 4}) (\delta_{\beta, 3} + \delta_{\beta, 4})}{\delta E_-^{(2)}} &
\frac{\xi^2_2(\delta_{\alpha, 3} - \delta_{\alpha, 4}) (\delta_{\beta, 1} - \delta_{\beta, 2})}{\delta E_+^{(2)}}  - \frac{(\delta_{\alpha, 1} + \delta_{\alpha, 2}) (\delta_{\beta, 3} + \delta_{\beta, 4})}{\delta E_-^{(2)}} \\
\frac{\xi^2_2(\delta_{\alpha, 1} - \delta_{\alpha, 2}) (\delta_{\beta, 3} - \delta_{\beta, 4})}{\delta E_+^{(2)}}  - \frac{(\delta_{\alpha, 3} + \delta_{\alpha, 4}) (\delta_{\beta, 1} + \delta_{\beta, 2})}{\delta E_-^{(2)}} &
\frac{\xi^2_2(\delta_{\alpha, 3} - \delta_{\alpha, 4}) (\delta_{\beta, 3} - \delta_{\beta, 4})}{\delta E_+^{(2)}}  +\frac{(\delta_{\alpha, 1} + \delta_{\alpha, 2}) (\delta_{\beta, 1} + \delta_{\beta, 2})}{\delta E_-^{(2)}}  \\
\end{pmatrix},
\end{equation}
\end{widetext}
where these $2\times 2$ matrices act on the effective spin 1/2 degree of freedom constituted by the odd ground states.

The Hamiltonian \eqref{hkondo} with the previous matrices  allows us to define the Kondo problem associated to the poor man's tetron as a function of the physical parameters of the system. For the sake of simplicity, in the next subsections we will restrict to the 3-lead problem by setting $J_4=0$.

\subsection{The poor man's topological Kondo point} \label{PMTK}

Topologically protected tetrons with four Majorana modes, are expected to display a topological Kondo effect characterized by a fractional off-diagonal conductance $2e^2/Mh$ where $M=3,4$ is the number of external leads coupled to the zero-energy modes \cite{Altland2013,Beri2012,Beri2013}. This conductance is the main signature of the topological Kondo fixed point, which is stable against anisotropies of the couplings with the leads \cite{Beri2012}, but it is unstable under the Majorana overlaps, that constitute an effective relevant Zeeman term in the related Kondo problem \cite{Altland2014,Altland2014b}. Moreover, renormalization group analysis and numerical simulations of the dynamics of tetrons show that the topological Kondo effect is magnified in proximity of the charge degeneracy point of these devices \cite{Vayrynen_PRR2020,Wauters_PRB2023}. 

In the following we investigate under which tuning conditions the poor man's tetron displays an analogous topological Kondo point that can be detected by a fractionalized conductance at low voltage bias. 

As showed by the definition of the effective Hamiltonian in Eq. \eqref{hkondo} and the matrices $\Upsilon$ and $\Xi$, the Kondo problem emerging from the poor man's tetron depends on the energies of the eigenstates of the Hamiltonian \eqref{hameff}, which, in turn, are affected by the charging energy $E_C$. This is the main difference between the poor man's tetron and the topologically protected tetron and it originates from their different geometry entailing that, as a first approximation, the charging energy of the poor man's tetron does not depend on the occupation of the four external dots, but only on the charge $N$ of the superconducting island. This situation is different in the standard tetron in which the fermionic parity of the Majorana modes directly affects its charging energy.

To build the analogy between the Hamiltonian $H_{\rm Kondo}$ and the topological Kondo effect, we rewrite it in the following form:
\begin{equation} \label{hamkondo2}
H_{\rm Kondo} = \sum_{\alpha,\beta} J_\alpha J_\beta^{*} l^\dag_\alpha l_\beta \left( \Upsilon^{\alpha \beta} - \Xi^{\alpha \beta}\right) + \sum_\alpha \left|J_\alpha\right|^2 \Xi^{\alpha \alpha}\,.
\end{equation}
In this formulation, the leads are coupled to the two odd ground states via the matrices $\Upsilon^{\alpha \beta} - \Xi^{\alpha \beta}$, and the matrices $\Xi^{\alpha \alpha}$ may cause an additional Zeeman splitting of the effective spin 1/2 degree of freedom.
In the case with three leads, by following the construction in Refs.~\cite{Galpin2014,TemaismithiPhD}, we can define a set of spin-1 lead operators in the form:
\begin{equation} \label{generators}
I_\gamma=i \sum_{\alpha,\beta=1}^3 \epsilon_{\gamma \beta \alpha} l^\dag_\alpha l_\beta\,,
\end{equation}
where $\epsilon$ is the Levi-Civita antisymmetric tensor. These operators correspond to generators of the $SO(3)$ group that underlies the topological Kondo effect.

A topological Kondo point is indeed achieved when $H_{\rm Kondo}$ acquires the ideal form \cite{Beri2012,Galpin2014}:
\begin{equation} \label{hamTK}
H_{\rm TK} = \sum_\gamma \lambda_{\gamma \gamma} I_\gamma \sigma_\gamma \,,
\end{equation}
which constitutes an overscreened Kondo problem with the spin-1 lead operators $I_\alpha$ coupled to the effective spin 1/2 Pauli matrices $\sigma_\gamma$. Additional Zeeman terms affecting the spin 1/2 degree of freedom are relevant perturbations that must vanish to retrieve the ideal topological Kondo effect.

To compare the Hamiltonians \eqref{hamkondo2} and \eqref{hamTK}, we define the matrices
\begin{equation}
\Sigma^{\alpha\beta} \equiv \Upsilon^{\alpha\beta} - \Xi^{\alpha \beta} \,,
\end{equation}
and we consider their symmetrization and antisymmetrization under the exchange $\alpha \leftrightarrow \beta$:
\begin{equation}
\Sigma^{\alpha\beta}_{\rm s} \equiv \frac{\Sigma^{\alpha\beta} + \Sigma^{\beta\alpha}}{2} \,,\qquad
\Sigma^{\alpha\beta}_{\rm as} \equiv \frac{\Sigma^{\alpha\beta} - \Sigma^{\beta\alpha}}{2}\,.
\end{equation}
Introducing the constants
\begin{align}
&A_{\pm} = \frac{\cos^2\theta \xi_4^2}{2\delta E_+^{(4)}} \pm \frac{\sin^2\theta \xi^2_2}{2 \delta E_+^{(2)}}\,, \label{Aplus}\\
&B_{\pm} = \frac{\cos^2\theta }{2\delta E_-^{(4)}} \pm \frac{\sin^2\theta }{2 \delta E_-^{(2)}} \label{Bplus}\,,
\end{align}
these may be expressed as
\begin{align}
&\Sigma^{12}_{\rm s} = - \sigma_3 \frac{A_+ + B_+}{2}  + \sigma_0 \frac{B_+-A_+}{2}\,,\quad \Sigma^{12}_{\rm as}=0\,,\\
&\Sigma^{23}_{\rm s} = - \sigma_1 \frac{A_+ + B_+}{2}  \,, \qquad\Sigma^{23}_{\rm as}=  i\sigma_2 \frac{A_+-B_+}{2}\,,\\
&\Sigma^{13}_{\rm s} = \sigma_1\frac{B_- - A_-}{2}  \,, \qquad \;\;\,\Sigma^{13}_{\rm as}=  i\sigma_2 \frac{A_-+B_-}{2}\,.
\end{align}
Without loss of generality, we consider a gauge choice for the leads degrees of freedom such that $J_2=J$ is real and $J_1=J_3=iJ$ are imaginary. Based on these definitions, the Hamiltonian \eqref{hamkondo2} can be rewritten as:
\begin{multline} \label{hamkondo3}
\frac{H_{\rm Kondo}}{J^2} = -I_3 \Sigma^{12}_{\rm s} - I_1 \Sigma^{23}_{\rm s} -iI_2 \Sigma^{13}_{\rm as}\\
+iF_1 \Sigma^{23}_{\rm as} + F_2 \Sigma^{13}_{\rm s} \\ +u Z \sigma_3
+\sum_\alpha \Xi_{\alpha \alpha}-w \sum_\alpha l^\dag_\alpha l_\alpha\,,
\end{multline}
where $u=\left(B_--A_-\right)/2$, $w=\left(A_-+B_-\right)/2$, and we introduced the following (symmetric) lead operators:
\begin{align} \label{sympert}
& F_\gamma = \sum_{\alpha,\beta=1}^3 \left|\epsilon_{\gamma \alpha \beta} \right|l^\dag_\alpha l_\beta\,,\\
& Z=l^\dag_1l_1+ l^\dag_2 l_2 - l^\dag_3 l_3 \,.
\end{align}
The operators $F$ correspond to generators of the $SU(3)$ group, whereas the operator $Z$ is not traceless and can be decomposed in terms of generators of $U(3)$.

In Eq. \eqref{hamkondo3} the last term plays the role of a diagonal potential scattering and it can be simply considered as a shift of the voltage bias of the leads, therefore we disregard it. 

The most straightforward approach to a topological Kondo point is to set the constraint $A_+=B_+$:
\begin{equation} \label{tuningTK}
\frac{\cos^2\theta \xi_4^2}{2\delta E_+^{(4)}} + \frac{\sin^2\theta \xi^2_2}{2 \delta E_+^{(2)}} =\frac{\cos^2\theta }{2\delta E_-^{(4)}} + \frac{\sin^2\theta }{2 \delta E_-^{(2)}}\,.
\end{equation}
Under this assumption, we obtain indeed:
\begin{multline} \label{hamkondof}
H_{\rm Kondo} = J^2 \left(A_+I_1\sigma_1 + \frac{A_- + B_-}{2} I_2 \sigma_2 + A_+ I_3\sigma_3  \right) \\
+ J^2\frac{B_--A_-}{2}\left(F_2\sigma_1 +
Z\sigma_3 - \frac{\sigma_3}{2}
\right)\,. 
\end{multline}
Here, the first line matches the description of the 3-lead topological Kondo Hamiltonian \eqref{hamTK}, with an anisotropic coupling along the $y$ direction of the spins; the topological Kondo effect, however, is robust against this kind of anisotropy \cite{Beri2012}. The second line of equation \eqref{hamkondof}, instead, corresponds to perturbations away from the topological Kondo fixed point, which share the same coupling:
\begin{equation} \label{pertamplitude}
\varepsilon \equiv  J^2\frac{B_--A_-}{2} = \frac{J^2\cos^2\theta}{2}\left(\frac{1}{\delta E_-^{(4)}}-\frac{\xi_4^2}{\delta E_+^{(4)}}\right)
\end{equation}
at the fine-tuned point provided by Eq. \eqref{tuningTK}. We observe that $\varepsilon \to 0$ in the limit of small charging energies.

\begin{figure}
\includegraphics[width=\columnwidth]{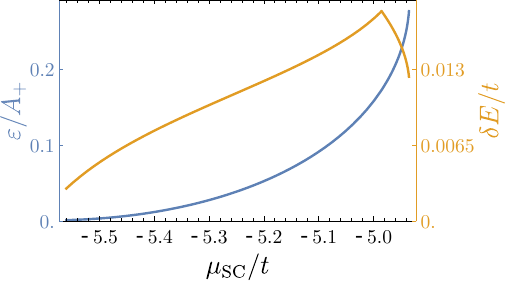}
\caption{Relative error $\varepsilon/A_+$ (blue) and energy gap $\delta E$ (orange) between the even excited states and the odd ground states as a function of $\mu_{\rm SC}$. The data are calculated by considering the optimal induced charge $n_g$ fulfilling the constraint \eqref{tuningTK}. The other system parameters are the same as Fig.~\ref{fig:spectrum}.}
\label{fig:error}
\end{figure}

For the parameters adopted in Fig.~\ref{fig:spectrum}, the condition \eqref{tuningTK} is fulfilled for $n_g \approx 0.997$ where the ratio between the perturbation \eqref{pertamplitude} and the coefficient $\lambda_{11}$ of the topological Kondo Hamiltonian becomes $\varepsilon / A_+ \approx 0.15$. In Fig.~\ref{fig:error} we observe, however, that by varying the parameter $\mu_{\rm SC}$ it is possible to achieve considerably lower relative perturbations $\varepsilon / A_+$ at the fine-tuned points fulfilling Eq. \eqref{tuningTK} in correspondence of sizeable gaps between the odd ground states and the even excited states. 

Requiring that both the tuning \eqref{tuningTK} is fulfilled and the perturbation $\varepsilon$ vanishes corresponds to setting:
\begin{equation}
\xi^2_{N_t} = \frac{\delta E_+^{(N_t)}}{\delta E_-^{(N_t)}}
\end{equation}
for both $N_t=2,4$. This double condition cannot in general be achieved by tuning only $n_g$, but it requires a simultaneous control of at least two parameters (for instance $n_g$ and $\mu_{\rm SC}$).

Importantly, the constraint \eqref{tuningTK} does not depend on the amplitudes of the couplings $|J_\alpha|$. Anisotropic couplings $J_\alpha$ would simply modify the different terms in Eqs. \eqref{hamkondo3} and \eqref{hamkondof} through multiplicative constants. Since the topological Kondo effect is robust against the anisotropy of the coefficients $\lambda$ in Eq. \eqref{hamTK}, small differences in the parameters $J_\alpha$ do not hinder the possibility of approaching the topological Kondo point.

In the following, we show that the $\varepsilon$ perturbations in the second line of the Hamiltonian \eqref{hamkondof} are relevant in the renormalization group sense and we discuss their effect.

\subsection{Effects of perturbations} \label{sec:RG}

The Kondo Hamiltonian in Eq. \eqref{hamkondof} differs from the topological Kondo Hamiltonian in Eq.~\eqref{hamTK} by the additional exchange interactions $F_2 \sigma_1$ and $Z\sigma_3$, besides a Zeeman term along $\sigma_3$. Furthermore, in realistic scenarios, we must consider several additional effects that concur to perturb the topological Kondo point identified by the optimal tuning \eqref{tuningTK}. 

The first is determined by the contribution of the excited states of the even sectors $(N_t,-)$ with $N_t=2,4$ in the perturbative definition of the matrices $\Upsilon$ and $\Xi$ (see Appendix \ref{app:exci}). These high-energy states cause a shift of the optimal point [as described in Eq. \eqref{TKtuning}], modify the Zeeman field along $\sigma_3$ and introduce additional potential scattering terms proportional to $l^\dag_2 l_1 \sigma_0 + {\rm H.c.}$. 

The second important perturbation is caused by a possible detuning away from the optimal condition in Eq. \eqref{tuningTK} (or, analogously, Eq. \eqref{TKtuning}), which determines the onset of the term $F_1 \sigma_2$ in Eq. \eqref{hamkondo3}.

Therefore, in general we will consider the following Hamiltonian for a 3-lead setup:
\begin{equation}
H=H_{\rm TK} + \delta H\,,
\end{equation}
where the topological Kondo Hamiltonian $H_{\rm TK}$ is defined in \eqref{hamTK}, and the perturbations away from it acquire the general form:
\begin{multline} \label{corrections}
\delta H = h_3 \sigma_3 + \varepsilon_2 \left( l^\dag_1 l_3 \sigma_1 + \rm{H.c}\right) + \varepsilon_1 \left( l^\dag_2 l_3 \sigma_2 + \rm{H.c}\right) \\ +\varepsilon_2 \left(l^\dag_1l_1+l^\dag_2l_2-l^\dag_3 l_3\right)\sigma_3+ \sum_{\alpha,\beta} w_{\alpha \beta} l^\dag_\alpha l_\beta \,,
\end{multline}
with $\varepsilon_2 \propto J^2(B_- - A_-)$ and $\varepsilon_1 \propto J^2(B_+-A_+)$ corresponding to the detuning away from the condition \eqref{tuningTK}.

In the absence of the corrections $\delta H$, a device described by $H_{\rm TK}$ displays the topological Kondo effect below the Kondo temperature $T_K \propto \rho \lambda^2\exp[-1/(\rho \lambda)]$ that depends on the lead density of states $\rho$ and the coupling constants $\lambda_{\gamma \gamma}$ ($\lambda=\lambda_{\gamma\gamma}$ in the isotropic limit \cite{Galpin2014}). $T_K$ dictates the crossover between this strong-coupling regime and weak-coupling regimes that can typically be addressed with resonant level models \cite{Vayrynen_PRR2020,Wauters_PRB2023}. The strongly coupled topological Kondo regime, in particular, is characterized by the onset of the fractional non-local conductance $2e^2/Mh$ whose corrections with the voltage bias scale with exponents reflecting its non-Fermi liquid behavior \cite{Zazunov2014,Beri2017} (see also \cite{Buccheri2022} for the related thermal transport properties).

When including the perturbations described by $\delta H$, a general analysis of their effects can be obtained by renormalization group considerations.
The last term in Eq. \eqref{corrections} describes a potential scattering that does not involve the effective spin 1/2 degree of freedom. This term is marginal in the renormalization group analysis of the topological Kondo problem determined by $H_{\rm TK}$. As previously mentioned, the components with $\alpha=\beta$ can be corrected by a suitable shift of the voltage bias of the leads. The off-diagonal terms with $\alpha \neq \beta$ do not qualitatively affect the non-Fermi liquid properties of the topological Kondo effect \cite{Galpin2014}. 

The Zeeman term $h_3 \sigma_3$ is relevant and drives the system away from the topological Kondo point if $h_3 \gtrsim T_K$. A small Zeeman term introduces an additional energy scale $T_h \sim T_K \left(h_3/T_K \right)^{3/2}$ \cite{Altland2014,Altland2014b}. $T_h$ sets the crossover between the topological Kondo effect with $M=3$ leads for $T_h \ll T \ll T_K$ and a trivial regime for $T\ll T_h$ with vanishing zero-bias conductance. 
Hence, in order to observe the physics of the topological Kondo effect, the Zeeman term $h_3$, as estimated in \eqref{pertamplitude}, must be sufficiently smaller than $T_K$. 

To investigate the role of the perturbations $\varepsilon_i$ in three-lead devices, we follow the construction in Ref.~\cite{Galpin2014} and introduce a set of rotated creation and annihilation lead operators of the form:
\begin{equation} \label{eq:ell}
\ell_{+1} = \frac{l_1+il_2}{\sqrt{2}}\,,\quad \ell_{-1} = \frac{-l_1+il_2}{\sqrt{2}}\,,\quad \ell_{0}= l_3\,.
\end{equation}
The Hamiltonian $H_{\rm Kondo}'= H_{\rm TK} + \delta H'$ can be rewritten as:
\begin{subequations} \label{hamgalpin}
\begin{align} 
& \hspace{-1em} H_{\rm Kondo}'= \nonumber\\
&\frac{\lambda_{11}+\lambda_{22}+\varepsilon_2-\varepsilon_1}{\sqrt{2}} \left(\ell^\dag_{0} \ell_{+1} \sigma^{+} + \ell^\dag_{+1} \ell_{0} \sigma^{-} \right) + \label{and1}
\\
&\frac{\lambda_{11}+\lambda_{22}-\varepsilon_2+\varepsilon_1}{\sqrt{2}} \left(\ell^\dag_{0} \ell_{-1} \sigma^{-} + \ell^\dag_{-1} \ell_{0} \sigma^{+} \right) + \label{and2}\\
&\frac{\lambda_{11}-\lambda_{22}+\varepsilon_2+\varepsilon_1}{\sqrt{2}} \left(\ell^\dag_{0} \ell_{+1} \sigma^{-} + \ell^\dag_{+1} \ell_{0} \sigma^{+} \right) +\label{and3}\\
&\frac{\lambda_{11}-\lambda_{22}-\varepsilon_2-\varepsilon_1}{\sqrt{2}} \left(\ell^\dag_{0} \ell_{-1} \sigma^{+} + \ell^\dag_{-1} \ell_{0} \sigma^{-} \right) + \label{and4}\\
& \left(\lambda_{33} +\varepsilon_2\right) \left(\ell^\dag_{+1}\ell_{+1}- \ell^\dag_{0}\ell_{0} \right)\sigma_3 \, +\label{and5}\\
&\left(\lambda_{33} -\varepsilon_2\right) \left(\ell^\dag_{0}\ell_{0}- \ell^\dag_{-1}\ell_{-1} \right)\sigma_3+\label{and6}\\
& + \varepsilon_2 \ell^\dag_0\ell_0\sigma_3 \label{and7}\,.
\end{align}
\end{subequations}
This form makes it easy to understand the peculiarity of the topological Kondo point and the role of the $\varepsilon_i$ perturbations.
In the unperturbed axial case defined by $\lambda_{11}=\lambda_{22}=\lambda$ and $\varepsilon_1=\varepsilon_2=0$, the three-channel topological Kondo problem in Eq. \eqref{hamTK} decomposes into two standard spin 1/2 Kondo problems with spin 1/2 conduction electrons. These two models, however, share the same impurity and one of their conduction channels, namely the one associated to the lead operators $\ell^\dag_0, \ell_0$ \cite{Galpin2014,TemaismithiPhD}. These two Kondo problems have their off-diagonal components defined by the first two lines in Eq. \eqref{hamgalpin}. For $\varepsilon_1=\varepsilon_2$, the two related amplitudes are equal and their competition is crucial to generate the non-Fermi liquid topological Kondo fixed-point that, indeed, corresponds to the overscreened Kondo problem with spin-1 conduction electrons. If $\varepsilon_2-\varepsilon_1 \neq 0$, however, one of the two Kondo problems dominates over the other. For instance, when $\varepsilon_2 > \varepsilon_1$, the first coupling constant in  Eq. \eqref{hamgalpin} grows faster than the second under the renormalization group and the system flows towards an isotropic strong coupling fixed point in which the channels $0$ and $+1$ screen the effective spin 1/2 impurity \cite{Galpin2014}. The perturbation corresponding to $\varepsilon_2-\varepsilon_1$ is thus relevant and drive the system from the topological Kondo point to a standard spin 1/2 Kondo problem.

The terms \eqref{and3} and \eqref{and4}, instead, correspond to the anisotropy of the couplings $\lambda_{\gamma \gamma}$ of the topological Kondo problem. This anisotropy is irrelevant \cite{Beri2012} which suggests that also the role of the perturbation associated to $\varepsilon_2 + \varepsilon_1$ does not affect the features of the topological Kondo point.

The terms \eqref{and5} and \eqref{and6} constitute the diagonal terms of the two spin 1/2 Anderson impurity models. Analogously to the related off-diagonal terms \eqref{and1} and \eqref{and2}, the perturbation $\varepsilon_2$ favors either of them, such that, also in this case, $\varepsilon_2$ appears as a relevant perturbation which strengthen the renormalization flow towards a fixed point in which the effective impurity is screened by two of the lead degrees of freedom represented by \eqref{eq:ell}, and either the $+1$ or $-1$ channel remains decoupled.

Finally, the last term \eqref{and7} may strengthen the role of the Zeeman term but does not lead to an asymmetry between the two standard Anderson models.

We conclude therefore that both perturbations $\varepsilon_2$ and $\varepsilon_1$, taken singularly, will drive the system away from the topological Kondo fixed-point. In particular, given the analogy to the axial anisotropy studied in Refs.~\cite{Galpin2014,TemaismithiPhD}, these perturbations behave in a way similar to the Zeeman terms and define an additional energy scale:
\begin{equation}
T_{\varepsilon} \sim T_K \left[\rho\left|\varepsilon_2-\varepsilon_1 \right|\right]^{3/2}\,.
\end{equation}
The fractional conductance and non-Fermi liquid features typical of the topological Kondo point are expected to emerge for bias voltages and temperatures in the interval $T_\varepsilon \ll T,eV_b \ll T_K$. If the perturbations $\varepsilon_i$ become comparable to the Kondo temperature $T_K$, instead, the topological Kondo features will not emerge in the poor man's tetron.

We end the analysis of the perturbations away from the topological Kondo point by considering the differences between poor man's tetrons coupled with $M=3$ and $M=4$ leads. In general, the Zeeman term in Eq. \eqref{hamkondo3} originates from the sum over the lead index of the operators $\Xi^{\alpha \alpha}$. In the four-lead case, when all the couplings $J_\alpha$ have the same modulus, this term simply returns a trivial contribution proportional to the $2\times 2$ identity in Eq. \eqref{hamkondof}. This makes it easier to observe the onset of the topological Kondo effect. 

The topological Kondo problem stemming from the four-lead devices can be analyzed by generalizing the lead operators $I_\gamma$ in Eq. \eqref{generators} to the antisymmetric lead generators of the $SO(4)$ group (see, for instance, \cite{Guanjie2023}). Its perturbations can be expressed in terms of symmetric lead generators of $SU(4)$ that generalize the $F_\gamma$ operators in Eqs. \eqref{sympert}. Furthermore, analogously to the Zeeman term, the $Z$ term in Eq. \eqref{hamkondo3} is replaced by additional $SU(4)$ generators in four-lead systems with equal couplings, such that, differently from the three-lead case, no perturbations driving the system towards a $U(4)$ Kondo model appear.

The terms stemming from symmetric generators are relevant perturbations also for $M=4$, but the related crossover energy between trivial and topological Kondo regimes is expected to scale as $\varepsilon^2/T_K$. Therefore, we envisage that the poor man's topological Kondo regime is more easily observable in four-lead devices. A perturbative renormalization group analysis of this problem is deferred to future works.
 
\subsection{The Anderson impurity model regime} \label{sec:anderson}

The decomposition of the perturbed topological Kondo Hamiltonian in the form \eqref{hamgalpin} suggests that driving the system away from the topological Kondo point defined by the constraint \eqref{tuningTK} modifies the low-temperature behavior of the poor man's tetron from the non-Fermi liquid topological Kondo fixed point to a strongly coupled Kondo regime corresponding to an effective spin 1/2 impurity screened by two species of conduction electrons.

This picture can be further confirmed by the analysis of the Kondo problem in proximity to the degeneracy points $n_g = n_g^{(1/2)}$ (see Fig.~\ref{fig:spectrum}). In these points, one of the even ground states becomes degenerate with the odd ground states. In the following we show that such regime can be mapped onto a simple spin 1/2 Anderson impurity model with infinite interaction, which, indeed, shares the low-temperature behavior of the standard Kondo problem.

We focus in particular on the neighborhood of $n_g \approx n_g^{(1)}$. The behavior at $n_g^{(2)}$ is analogous through a suitable particle-hole transformation.

Close to $n_g^{(1)}$ the lowest energy states of the poor man's tetron are the almost-degenerate $\ket{3+}, \ket{3-}$ and $\ket{2+}$. The lowest excitation is then provided by $\ket{2-}$, which is separated in energy by $\delta E_-^{(2)} \sim \sqrt{2} \Delta_{\rm CAR} - 2 \left|t_{\rm COT}(0)\right|$. For different parameter regimes with respect to the one depicted in Fig.~\ref{fig:spectrum} the roles of $\ket{2+}$ and $\ket{2-}$ may be exchanged.

We consider a regime in which the $J_\alpha$ couplings with the leads in Eq. \eqref{dotcoupl} are small compared to the energy gaps of the tetron: $|J_\alpha| \ll \delta E_-^{(2)}$. Therefore we can approximate the low-energy dynamics of the system by considering only the three ground states.
Given the fact that $\ket{2+}$ has one electron less than the odd states, the resulting effective Hamiltonian acquires the form:
\begin{multline} \label{spinpolAIM}
H\left(n_{g}^{(1)}\right)= \\
-\sum_{\alpha,b} J_\alpha l_\alpha^\dag \bra{2+} d_\alpha \ket{3b} \, \ket{2+}\bra{3b} + {\rm H.c.}\\
= \frac{\xi_2\sin\theta}{\sqrt{2}} \left[\left(J_2l_2^\dag - J_1l_1^\dag\right) \ket{2+}\bra{3+} \right. \\ \left. + \left(J_4l_4^\dag - J_3l_3^\dag\right) \ket{2+}\bra{3-}\right] + {\rm H.c.}\,,
\end{multline}
where we used Eq. \eqref{dm}.
From the previous equation it is easy to observe that only two linear combinations of the lead operators are involved in the Hamiltonian. In particular, we relabel:
\begin{equation}
\ell_\Upt \propto  J_2^*l_2 - J_1^*l_1\,,\qquad \ell_\Dnt \propto J_4^*l_4 - J_3^*l_3\,.
\end{equation}
The Hamiltonian $H\left(n_{g}^{(1)}\right)$ becomes equivalent to the spin 1/2 Anderson impurity model:
\begin{equation} \label{HamAIM}
H_{\rm AIM} = \left(J_\Upt \ell_\Upt^\dag d_\Upt + J_\Dnt \ell_\Dnt^\dag d_\Dnt + {\rm H.c. }\right) + U d^\dag_\Upt d_\Upt d^\dag_\Dnt d_\Dnt\,,
\end{equation}
in the limit of infinite interactions $U\to \infty$. This mapping corresponds to considering the even state $\ket{2-}$ as the vacuum state of the impurity problem, and the odd states as the doublet defined by the creation operators $d^\dag_\Upt$ and $d^\dag_\Dnt$.

In this scenario, there is no term in Hamiltonian \eqref{HamAIM} that allows for a flip of the $\tau$ degrees of freedom of the electrons, such that any differential conductance between the top and bottom wires vanishes at zero bias. This constitutes the main distinction in the transport features between the topological Kondo regime discussed in Sec. \ref{PMTK} and the Anderson model regime with strong $\varepsilon$ perturbations.
Other differences between these two regimes affect the transport from the left to the right leads: we observe that the model in Eq. \eqref{spinpolAIM} displays strong similarities with the infinite interaction limit of a two-channel impurity model with spin polarized leads \cite{Martinek2003,Paaske2008}, with the $\tau$ degree of freedom playing the role of the spin. For anisotropic amplitudes $J_\alpha$ that break the symmetry between the two nanowires, it is known that the Kondo peak of this model is split by a different renormalization of the doublet energy levels. Consequently, the differential conductance of the poor man's tetron between left and right leads within the same nanowire is expected to exhibit a splitting of the Kondo peak away from zero bias when the $\varepsilon$ perturbations dominate.

\section{Numerical results beyond the perturbative regime} \label{sec:surrogate}

\begin{figure*}[th!] 
\includegraphics[width=0.9\textwidth]{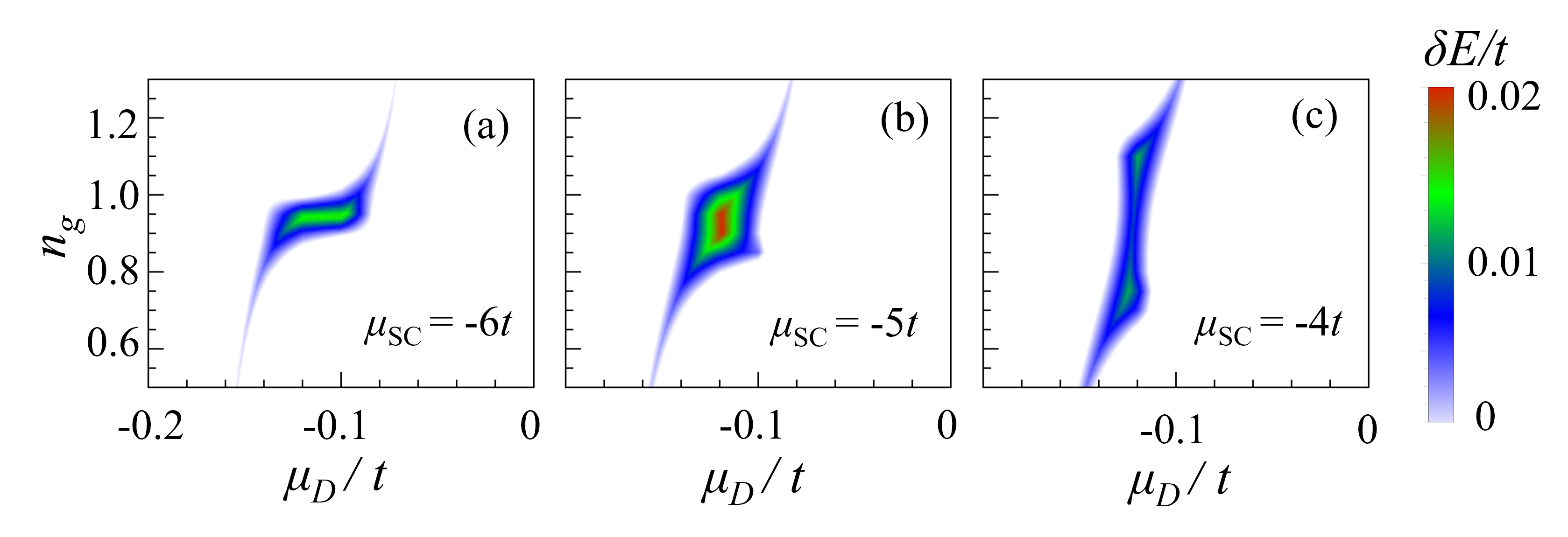}
\caption{Excitation energy $\delta E/t=\min\left(\delta E^{(2)}_\pm,\delta E^{(4)}_\pm \right)/t$ vs $
\mu_D$ and $
n_g$ for various $\mu_{\text{SC}}/t=-6,-5,-4$. Only the region with the odd $N_t=3$ ground state is shown. The other parameters are $\Delta=4t$, $\alpha=\pi/6$, $E_C=0.15t$, $E_Z=1.5\Delta$. }
\label{fig:numeric_1}
\end{figure*}

The analysis performed in the previous sections relied on two main simplifying assumptions: (i) we assumed that the dots were weakly coupled to the superconducting islands, namely that $t \ll \tilde{\Delta}-E_C$, to describe the dynamics in terms of CAR and coherent cotunneling processes; (ii) we considered spin polarized systems by assuming that the electrons in the dots were affected by a strong Zeeman term $\left(gB_z \gg \bar{t}_{\rm COT}, \Delta_{\rm CAR} \right)$ and possibly by strong intradot repulsions.

The poor man's tetron, however, can be analyzed in more general regimes by performing exact diagonalization on the full Hamiltonian $H_{\rm sys}= H_{\rm sp} + H_{\rm c}$ defined by Eqs. \eqref{hamtot} and \eqref{charging}, which we implement numerically using the Sneg software \cite{Zitko2011Oct,sneg2}.  In this section, we observe that the onset of ground states with odd fermionic parity can actually be favoured when increasing the tunneling amplitude $t$ beyond the validity of our previous perturbative analysis.

Additionally, since the energy splitting between the ground states of the different parity sectors mostly scales with the tunneling amplitude $t$, the non-perturbative regimes with a stronger hybridization between dots and superconducting island offers larger energy gaps. This is analogous to the enhanced stability of poor man's Majorana modes experimentally observed in the strong coupling regime \cite{Zatelli2023}.

Fig.~\ref{fig:numeric_1} shows how the parameter region with degenerate $N_t=3$ ground state in the $\mu_{\rm D},n_g$ plane evolves for different values of the potential $\mu_{\rm SC}$ in the weak tunneling regime. The related energy gap $\delta E$ is calculated in units of $t$ and the numerical data correspond to a strong polarized scenario in which Zeeman term $gB_z=1.5 \Delta=6t$. Fig.~\ref{fig:numeric_1}(b) corresponds to the parameters adopted in Fig.~\ref{fig:sectors}. To make the comparison easier we shifted $\mu_{\rm D}$ in Fig.~\ref{fig:numeric_1} by $gB_z$. See also panel (a) in Fig.~\ref{fig:numeric_3}. With respect to Fig.~$\ref{fig:sectors}$ we observe that $\mu_D$ and $\mu$ are displaced by a quantity roughly corresponding the average contribution $-\bar{t}_{\rm COT}/\cos 2\alpha = -2\bar{t}_{\rm COT}$ in Eq. \eqref{hameff}. Besides this, the exact diagonalization results in Fig.~\ref{fig:numeric_1} confirm the general validity of the perturbative results in Sec. \ref{PMtetron}. 

\begin{figure*}[th!] 
\includegraphics[width=0.9\textwidth]{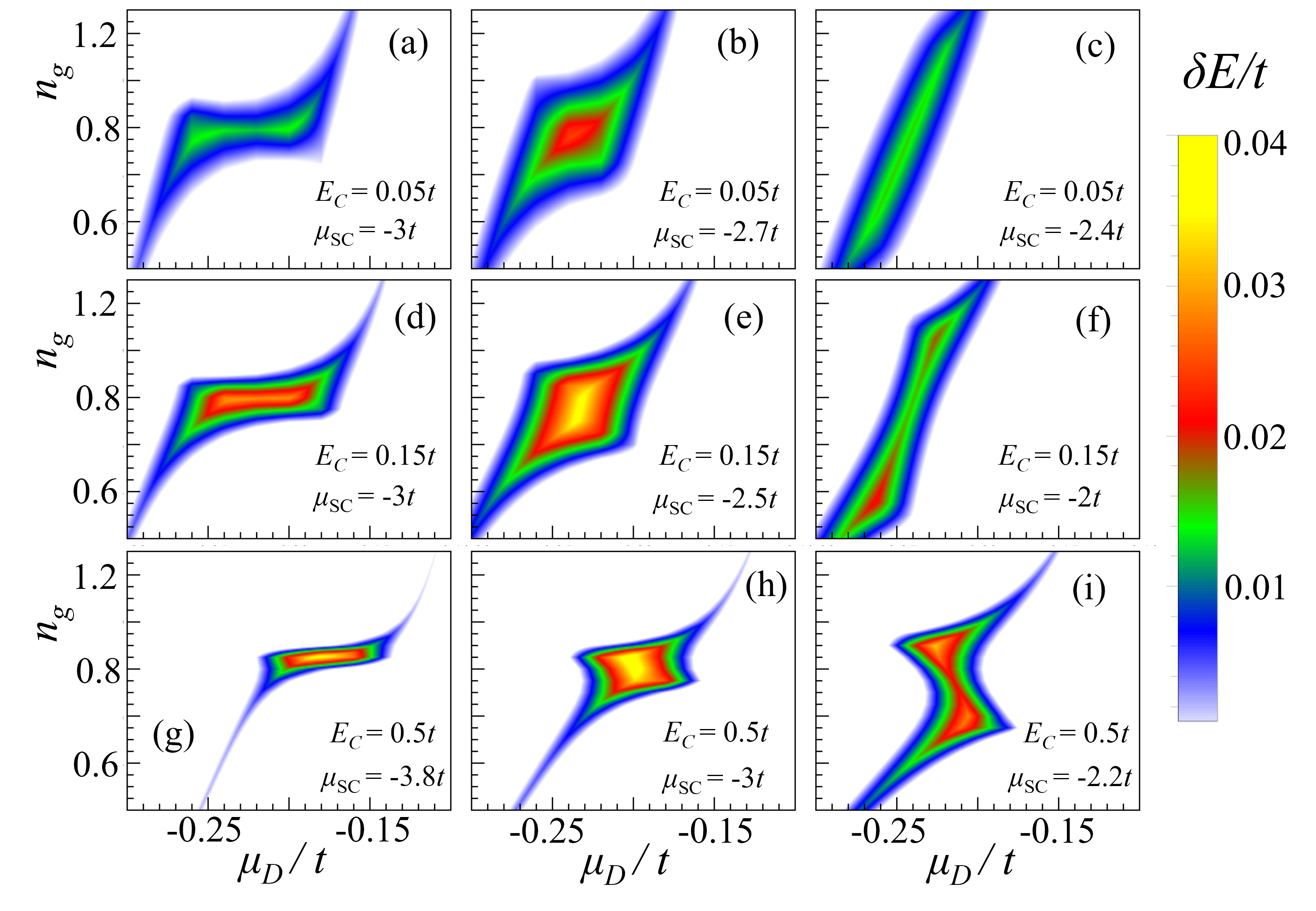}
\caption{Excitation energy $\delta E/t=\min\left(\delta E^{(2)}_\pm,\delta E^{(4)}_\pm \right)/t$ vs $
\mu_D$ and $
n_g$ in the non-perturbative regime at $\Delta=2t$ for various $\mu_\text{SC}$ and  charging energies. Only the region with the odd $N_t=3$ ground state is shown. The other parameters are $\alpha=\pi/6$, $E_Z=1.5\Delta$. The color scale and the ranges of the parameter $n_g$ are the same as in Fig.~\ref{fig:numeric_1}.}
\label{fig:numeric_2}
\end{figure*}

Fig.~\ref{fig:numeric_2} shows instead the results in a regime with a stronger hybridization between the dots and the superconducting island. In particular we considered $\Delta=2t$ and we illustrate the extension of the region with degenerate $N_t=3$ ground states in the $\mu_{\rm D},n_g$ plane for different values of $\mu_{\rm SC}$. In this scenario, the odd ground states dominate over a considerably broader region in parameter space and the related gaps grow with the tunneling amplitude $t$. The comparison between Fig.~\ref{fig:numeric_2} and Fig.~\ref{fig:numeric_1} suggests that the Kondo physics discussed in Sec. \ref{Kondo} is more easily observable for large tunnel coupling $t$, where the tuning of the other parameters to obtain degenerate ground states is easier to achieve. The three panels in Fig.~\ref{fig:numeric_2} additionally display the extension of the region with $N_t=3$ degenerate ground states for different values of the charging energy ranging from a case in which $E_C$ is less than half of the corresponding CAR and cotunneling amplitudes to a strong $E_C$ example. We conclude that the degeneracy of the ground states of the poor man's tetron can be achieved for a broad range of charging energies.

\begin{figure}[th!]
\includegraphics[width=\columnwidth]{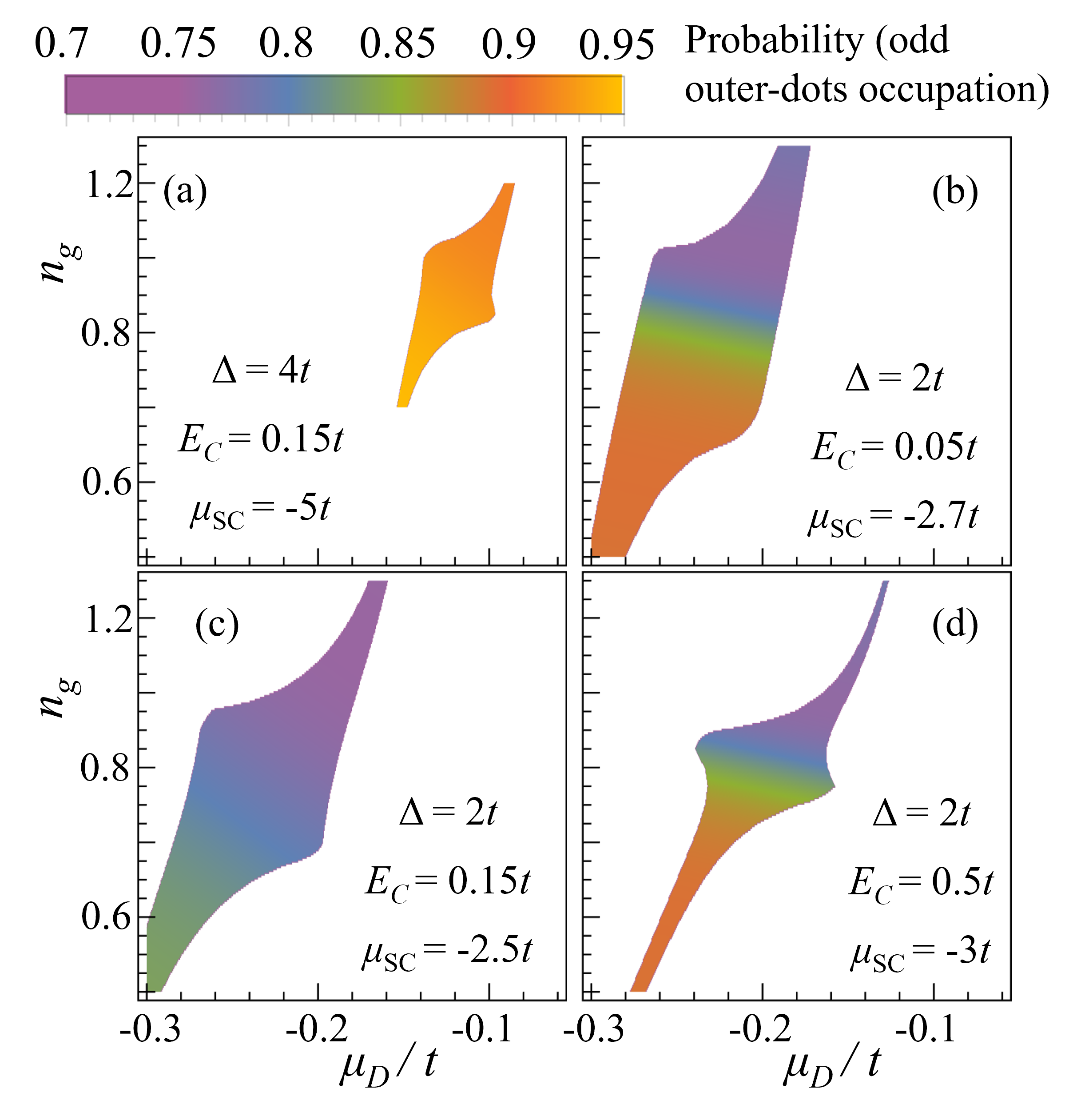}
\caption{Probability for odd occupation of the four quantum dots for the data displayed in Fig.~\ref{fig:numeric_1}(b) and panels (b), (e) and (h) of Fig.~\ref{fig:numeric_2}. The other parameters are $\alpha=\pi/6$, $E_Z=1.5\Delta$.}
\label{fig:numeric_3}
\end{figure}

Finally, we estimate the deviations of the exact odd ground states from the perturbative form presented in Eqs. \eqref{o1} and \eqref{o2}. The perturbative analysis in Sec. \ref{Kondo} is based on the assumption that the Bogoliubov states of the proximitized dots at energy $\tilde{\Delta}$ are not occupied. In this case the total fermionic parity matches with the fermionic parity of the external dots. Away from the perturbative regime, however, this assumption is not fulfilled in general, such that ground states with odd $N_t$ present also components with an even fermionic parity of the total occupation $\sum_{a,s,\tau} d^\dag_{as\tau} d_{as\tau}$. These components do not have a counterpart in the description given in Eqs. \eqref{o1} and \eqref{o2} and may cause significant deviations from the Kondo problem analyzed in Sec. \ref{Kondo}. In Fig.~\ref{fig:numeric_3} we depict the probability of observing a total odd fermionic parity of the external dots in the exact ground states with $N_t=3$. In the perturbative regime [panel (a)] this probability is always above $0.9$ such that we do not expect significant non-perturbative errors in the analysis of Sec. \ref{Kondo}. When increasing $t$ [panels (b,c,d)] this probability may decrease for certain choices of the parameters. However, even at $t/\Delta =0.5$ we typically observe that this probability is beyond $0.7$ such that the exact ground states do not excessively deviate from the analyzed $\ket{3+}$ and $\ket{3-}$ and we do not expect major deviations from the Kondo models analyzed in Sec. \ref{Kondo}.  

\section{Conclusions}

The experimental realization of minimal Kitaev chains in single-nanowire devices \cite{Dvir2023,Bordin2023,Zatelli2023,bordin2024} provides a platform to study key properties of Majorana modes by trading off their topological protection for an improved controllability of these systems.

The poor man's Majorana tetron that we presented constitutes a fundamental element to extend the study of the interplay between elastic cotunneling and crossed Andreev reflection processes to interacting hybrid systems.

We proposed a blueprint for this device, which is composed by four quantum dots coupled via a floating superconducting island, and we characterized its low-energy features. The tuning of its physical parameters enables the observation of the topological Kondo effect emerging in multiterminal conductance experiments when the poor man's tetron is coupled to external leads.

The physics of the related four poor man's Majorana modes is heavily altered by the introduction of the charging energy in the poor man's tetron: this device is characterized by local electrostatic interactions that, in general, separately affects the dots and the superconducting island mediating the interactions among them. These interactions are therefore qualitatively different from the topologically protected tetrons \cite{Karzig2017,Landau_PRL2016,Plugge2016,Plugge2017}, in which the fermionic parity encoded in the Majorana modes is directly linked with the charging energy of the Majorana-Cooper pair box. A similar distinction holds also for the comparison with hybrid quantum dot - superconducting island platforms for the observation of the topological symplectic Kondo effect \cite{Guanjie2023b}.

This difference in the form of the electrostatic interactions importantly affects both the structure of the many-body low-energy states of the poor man's tetron and their degeneracies, which in general are completely split by the charging energy even when tuning the poor man's Majorana modes to their sweet spot. Under this point of view, the poor man's tetron is also qualitatively different from its single nanowire counterpart which, instead, allows for a tuning of its parameters that enables the onset of zero-energy poor man's Majorana modes \cite{SoutoBaran}. 

Although poor man's Majorana modes are unstable against the charging energy in the poor man's tetron, we have shown that by carefully tuning the dot energy levels, the induced charge on the superconducting island, and the Andreev energy levels of the hybridized nanowires, extended regions of parameter space can be found where the ground state exhibits a two-fold degeneracy and behaves as an effective non-local spin-1/2 degree of freedom. This degeneracy is obtained if the system parameters are tuned such that a symmetry between the two nanowires emerges at low energies; in particular the respective crossed Andreev reflection and cotunneling amplitudes must match and their effective spin-orbit coupling must be sufficiently strong. When a degenerate configuration of the poor man's tetron is reached, the transport properties of this device become non-trivial and are characterized by the onset of an effective spin 1/2 Kondo impurity problem stemming from the different nanowire fermionic parities that distinguish the degenerate ground states.

We studied the resulting Kondo effect based on perturbation theory and we showed that under a suitable tuning of the system parameters, it approaches the topological Kondo regime that characterizes topologically protected tetrons. We examined the main perturbations that can drive the poor man's tetron away from the topological Kondo point in realistic devices and concluded that there exists a regime of parameters such that the fractional conductance of the topological Kondo point can be observed for intermediate temperatures and voltage biases. Since these perturbations are relevant in the renormalization group sense, however, the poor man's tetron in the low temperature and bias regime flows towards trivial fixed points corresponding to a suitably defined spin 1/2 Anderson impurity model. In this respect we confirmed that the topological Kondo effect \cite{Beri2012,Beri2013,Altland2013} in the limit of zero temperature and voltage bias is a genuine effect that reflects the topological protection of its constituent Majorana modes and it can be obtained by a poor man's Majorana device only through fine tuning.

We also discussed how our perturbative findings evolve in a non-perturbative regime through exact diagonalization and showed that a strong coupling between the dots and the superconducting island is beneficial to extend the the parameter region in which the poor man's tetron behaves as a two-level system.

We emphasize that in our analysis we modelled the coupling between the superconductor and the nanowires by considering simple intermediate dots with an effective mean-field proximity-induced superconducting pairing. A more accurate modelling of the induced superconductivity can be performed by the use of surrogate models \cite{Baran2023Dec,Baran2024Jun,Baran2023Apr} to include the role of the Bogoliubov quasiparticles in the system more realistically.

The poor man's tetron is a device with a rich structure of eigenstates whose hierarchy can be modified by tuning its external parameters.  Its realization is within reach both in double-nanowire devices \cite{Vekris2022}, and in 2D electron gases with gate-defined quantum dots. The latter scenario, for instance, can be achieved by doubling the architecture realized in Ref.~\cite{haaf2024} for the case of a 2-site Kitaev chain.

The electric tunability of this device will allow for extending the study of its multiterminal transport features beyond the analysis presented in this work, which primarily aimed at identifying regimes where the three-lead topological Kondo effect emerges as a recognizable phenomenon.

In particular, our analysis focused on setups with two-fold degenerate ground states with odd fermionic parity. We remark, however, that the regime we studied is not the only region in the physical parameters that displays degenerate ground states. Other configurations allow for the emergence of different effective spin degrees of freedom expected to favor the onset of a variety of Kondo effects and non-Fermi liquids. For instance, the poor man's tetron is a suitable platform to explore the physics of the charge Kondo effect emerging from transport phenomena mediated by Andreev processes and the related multi-channel Kondo effects \cite{Pustilnik2017}.

\section*{Acknowledgements}
We thank  A. M. Bozkurt, D. van Driel, C.-X. Liu, K. Flensberg, J. Vayrynen, M. Wauters, M. Wimmer and F. Zatelli for insightful discussions.
L.M. and M.B. acknowledge funding from the Villum Foundation (Research Grant No. 25310). L.M. has been supported by the research grant ``PARD 2023'', and ``Progetto di Eccellenza 23-27'' funded by the Department of Physics and Astronomy G. Galilei, University of Padua. V.V.B. is supported by a grant of the Romanian Ministry of Education and Research, Project No. 760122/31.07.2023 within PNRR-III-C9-2022-I9. This project was supported by the European Research Council (ERC) under the European Unions Horizon 2020 research and innovation programme under Grant Agreement No. 856526, the Swedish Research Council under Grant Agreement No. 2020-03412, and NanoLund

\appendix

\section{Perturbative definition of elastic cotunneling and crossed Andreev reflection} \label{app:pert}

In this appendix we present the details of the second-order perturbative calculation that defines elastic cotunneling and CAR processes in the poor man's tetron.
The following analysis extends the analogous description of devices with a single nanowire in Ref. \cite{SoutoBaran}. It applies when the tunneling amplitudes $t$ from the external dots to the central SC segments of the nanowires can be considered small with respect to the other energy scales in the setup. The unperturbed Hamiltonian accounts for the charging energy, the chemical potential of the dots and the energy of the Andreev subgap states in the SC segments. In particular, the single-particle Hamiltonian describing these subgap states corresponds to:
\begin{equation}
H_{SC} = \sum_{s,\tau} \tilde{\Delta}_s f^\dag_{s\tau} f_{s\tau}\,,
\end{equation}
with energies:
\begin{equation}
\tilde{\Delta}_{\Up/\Dn} = \sqrt{\mu_{\rm SC}^2 + \Delta^2} \pm g_{\rm SC} B_z\,. 
\end{equation}
The Bogoliubov modes are defined by
\begin{align} \label{Bog1}
&f_{\Up \tau} = u c_{\Up \tau}\ee^{-i\varphi/2} + v c^\dag_{\Dn \tau}\ee^{i\varphi/2}\,,\\
&f_{\Dn \tau} = u c_{\Dn \tau}\ee^{-i\varphi/2} - v c^\dag_{\Up \tau}\ee^{i\varphi/2}\,,\label{Bog2}
\end{align}
with:
\begin{equation}
u = \sqrt{\frac{1}{2}+\frac{\mu_{\rm SC}}{2\tilde{\Delta}}}\,,\quad v=\sqrt{\frac{1}{2}-\frac{\mu_{\rm SC}}{2\tilde{\Delta}}}\,,
\end{equation}
where $\tilde{\Delta}=\sqrt{\mu_{\rm SC}^2 + \Delta^2}$\,.
By inverting Eqs. \eqref{Bog1} and \eqref{Bog2} we get:
\begin{equation}
c^\dag_{s \tau} = \left( u f^\dag_{s \tau} -s v f_{\bar{s} \tau}\right)\ee^{-i\varphi/2} \,.
\end{equation}
In the last equation we consider $s=\pm 1$ for spin up and down respectively; $\bar{s}$ is a spin aligned opposite to $s$. Expressed in terms of the Bogoliubov modes $f_s$ the tunneling Hamiltonian reads:
\begin{multline}
H_t = \\
- t \sum_{\tau,a,s,s^{\prime}} \left[\left(u f^\dag_{s \tau} - s v f_{\bar{s} \tau}\right)
 \left(\ee^{-ia\alpha\sigma_y}\right)_{ss'}d_{\tau s' a}\ee^{-i \varphi/2} + {\rm H.c.}\right].
\end{multline}
Here we considered the same tunneling amplitude $t$ for all the dots. The operator $\ee^{-i\varphi/2}$ suitably increases the charge of the SC box. This operator is conjugate to $N$ and obeys the commutation relation $\left[N,\ee^{-i\varphi/2}\right]=\ee^{-i\varphi/2}$.

In the following, we also assume that the magnetic field $B_z$ is pointing up and it is strong enough to neglect the external dot states with the polarization $s=\Up$ ($gB_z \gg t$). Additionally, we assume that all the chemical potentials of the external dots are the same such that we consider only a single energy level $\mu$ for all the dot electrons with $s=\Dn$ spin.
Finally we consider a regime of strong induced pairing, such that $E_C<\tilde{\Delta} - g_{\rm SC} B_z$.
Under these assumptions, the elastic cotunneling at second order results:
\begin{widetext}
\begin{multline} \label{cot1}
H_{\rm COT} = \sum_{\tau, a^{\prime}, a^{\prime\prime}, s} -d^\dag_{a^{\prime \prime} \Dn \tau} \left(\ee^{i a^{\prime \prime} \alpha \sigma_y}\right)_{\Dn s} \frac{t^2u^2}{\tilde{\Delta}_s + H_{\rm c}(N_I+1) -  H_{\rm c}(N_I) -\mu}\left(\ee^{-i a^{\prime } \alpha \sigma_y}\right)_{s\Dn}d_{ a^{\prime } \Dn \tau} \\
-  d_{ a^{\prime \prime} \Dn \tau} \left(\ee^{-i a^{\prime \prime} \alpha \sigma_y}\right)_{s\Dn} \frac{t^2v^2}{\tilde{\Delta}_{\bar{s}} +  H_{\rm c}(N_I-1) -  H_{\rm c}(N_I)+\mu}\left(\ee^{i a^{\prime } \alpha \sigma_y}\right)_{\Dn s}d^\dag_{ a^{\prime } \Dn \tau}\,,
\end{multline}
\end{widetext}
where we labelled by $N=N_I$ the total charge of the superconducting island in the initial state. In the limit $E_C=0$ we recover the results in Ref.~\cite{Liu2022}.

In Sec. \ref{PMtetron}, we considered systems with $n_g \in [0,2]$ and, in this case, the lowest energy states of the unperturbed Hamiltonian $H_0$ correspond to the ones with charge $N=0$ and $N=2$ in the superconducting island. In particular, to have significant CAR processes, it is necessary to enforce the condition in Eq. \eqref{tuning} that guarantees the degeneracy between these states. Under this condition, the energy differences appearing in the denominators of Eq. \eqref{cot1} become:
\begin{align}
&H_{\rm c}(N_I+1) -  H_{\rm c}(N_I) -\mu = E_C\left(2N_I-1   \right)\,, \label{calc1}\\
&H_{\rm c}(N_I-1) -  H_{\rm c}(N_I)-\mu = E_C\left( 3 -2N_I\right)\,, \label{calc2}
\end{align}

The calculation of the relevant energy differences for different values of $N_I$ can be carried out straightforwardly.
The cotunneling Hamiltonian acquires the general form:
\begin{multline}
H_{\rm COT}(N_I) = \\
-\sum_{\tau, a^{\prime}, a^{\prime\prime}, s} d^\dag_{a^{\prime \prime} \Dn \tau} \left(\ee^{i a^{\prime \prime} \alpha \sigma_y}\right)_{\Dn s} \tilde{t}_s(N_I) \left(\ee^{-i a^{\prime } \alpha \sigma_y}\right)_{s\Dn}d_{ a^{\prime } \Dn \tau}\,,
\end{multline}
with the following values of the amplitudes in the most relevant sectors:
\begin{align}
\tilde{t}_s(2)&=t^2\frac{{\tilde{\Delta}_s u^2-\tilde{\Delta}_{\bar{s}} v^2}-E_C\left(u^2+3v^2\right)}{\left(\tilde{\Delta}_s+3E_C\right)\left({ \tilde{\Delta}_{\bar{s}}}-E_C\right)}\,,\\
\tilde{t}_s(0)&=t^2\frac{{\tilde{\Delta}_{\bar{s}} u^2-\tilde{\Delta}_s v^2}+E_C\left(3u^2+v^2\right)}{\left({\tilde{\Delta}_{\bar{s}}}+3E_C\right)\left(\tilde{\Delta}_s-E_C\right)}\,.
\end{align}
When considering $g_{\rm SC}=0$ and imposing $a^{\prime} \neq a^{\prime\prime}$, one obtains the cotunneling amplitudes in Eq. \eqref{cot}. The terms with $a^{\prime}= a^{\prime\prime}$, instead, correspond to the shift of the energy level $\mu$ we included in Eq. \eqref{hameff}.

The calculation of the CAR processes is performed in an analogous way. When considering transitions between states with $N_I$ and $N_I+2$ electrons in the superconducting island, we obtain an effective p-wave pairing between the dot degrees of freedom of the form:
\begin{widetext}
\begin{equation}\label{carH}
H_{\rm CAR}(N_I) = t^2 \ee^{-i\varphi} \sum_{\tau, a^{\prime}, a^{\prime\prime}, s} d_{a^{\prime \prime} \Dn \tau}\left(\ee^{-i a^{\prime \prime} \alpha \sigma_y}\right)_{\bar{s}\Dn} \frac{svu}{\tilde{\Delta}_s + H_c(N_I+1)  - {\left[ H_c(N_I) + H_c(N_I+2) \right]/2}}\left(\ee^{-i a^{\prime } \alpha \sigma_y}\right)_{s\Dn} d_{a^\prime \Dn \tau}  + {\rm H.c.}\,,
\end{equation}
\end{widetext}
where $s=\pm 1$ labels the spin of the virtual Bogoliubov state involved in the process.
Only processes between configurations with $N_I=0$ and $N_I=2$ connect degenerate states when we impose the condition \eqref{tuning}. However, the former expression for CAR processes provides an approximate value of the amplitudes when $\tilde{\Delta}_s \gg t \gtrsim E_C$.

Analogously to the cotunneling, we rewrite the CAR processes in the form: 
\begin{widetext}
\begin{equation}
H_{\rm CAR}(N_I) = \ee^{-i\varphi}
\sum_{\tau, a^{\prime}, a^{\prime\prime}, s} d_{a^{\prime \prime} \Dn \tau}\left(-i \sigma_y \ee^{i a^{\prime \prime} \alpha \sigma_y}\right)_{\Dn s} \tilde{\Delta}_{\rm CAR,s}(N_I) \left(\ee^{-i a^{\prime } \alpha \sigma_y}\right)_{s\Dn} d_{a^\prime \Dn \tau}  + {\rm H.c.}\,.
\end{equation}
\end{widetext}
The denominators in \eqref{carH} depend only on the curvature of the parabolas $H_c(N_I)$, hence the CAR amplitudes do not depend on the number of electrons $N_I$. We have:
\begin{equation}
\tilde{\Delta}_{\rm CAR,s}=\frac{t^2\sqrt{\tilde{\Delta}^2-\mu_{SC}^2}}{2\tilde{\Delta}\left(\tilde{\Delta}_s -E_C\right)}\,.
\end{equation}
For $g_{\rm SC}=0$, thus $\tilde{\Delta}_s = \tilde{\Delta}$, we obtain the result in Eqs. \eqref{hameff} and \eqref{CAR2}.

These calculations were performed under the simplifying assumption that the two nanowires are equivalent at the microscopic level. In realistic devices, however, we expect that the proximity induced pairing $\Delta$ and the effective Zeeman coupling $g_{\rm SC}$ acquire a dependence on $\tau$. 
Additionally, also the spin-orbit parameter $\alpha$ is subject to differences among the four dots.

In the general scenario, therefore, the effective cotunneling and CAR amplitudes are different for the two nanowires. This results, in turn, in a further dependence on $\tau$ of the amplitudes in Eqs. \eqref{hameff}, \eqref{cot} and \eqref{CAR2}.

In order to implement an emergent nanowire symmetry $\Upt \leftrightarrow \Dnt$ in the effective Hamiltonian \eqref{hameff} for any value of $n_g$, we need therefore to be able to separately control the ratios of the four amplitudes $t_{{\rm COT},\tau}$ and $\Delta_{{\rm CAR},\tau}$. The general ratio between amplitudes in the top and bottom nanowire can be tuned by varying the tunneling amplitudes $t_{a\tau}$ with suitable cutter gates. The remaining two ratios $t_{{\rm COT},\tau}/\Delta_{{\rm CAR},\tau}$, instead, require two additional controls to be suitably set.

In our previous analysis we assumed that the parameter $\mu_{\rm SC}$ is the same for both wires. However, by considering two separate side gate close to the superconducting island [labelled by $V_{g\Upt}$ and $V_{g\Dnt}$ in Fig. \ref{fig:tetron-schematic} (a)], also this potential acquires a dependence on $\tau$ such that the independent control of $\mu_{{\rm SC},\Upt}$ and $\mu_{{\rm SC},\Dnt}$ guarantees a sufficient tunability of the system to achieve the desired nanowire symmetry. Alternative strategies to tune the device can also be envisioned by considering that the cutter gates controlling the tunneling between the external dots and the central segments of the nanowires also affect the dot spin-orbit parameters.

\begin{widetext}
\section{Effective Hamiltonians in the four parity sectors} \label{app:sectors}
In this appendix, we report the $4 \times 4$ blocks of the effective Hamiltonian \eqref{hameff} under the conditions presented in Table \ref{table}.

\subsection{Odd sector $(N_t=2m+1,P_\Dnt=+)$ }

This is the sector with parities $P_\Upt=-1,\, P_\Dnt=1$.

Its basis is $\left\{\ket{0100,N_t-1},\ket{1000,N_t-1},\ket{0111,N_t-3},\ket{1011,N_t-3}\right\}$. The effective Hamiltonian for this sector reads:
\begin{equation}
H_{\rm eff}^{\text{odd}+} = \begin{pmatrix} 
\mu +H_c(N_t-1) & -t_{\rm COT}(N_t-1) & -\Delta_{\rm CAR} & 0 \\
-t_{\rm COT}(N_t-1) & \mu +H_c(N_t-1) & 0 & -\Delta_{\rm CAR} \\
-\Delta_{\rm CAR} & 0 & 3\mu + H_c(N_t-3) & -t_{\rm COT}(N_t-3) \\
0 & -\Delta_{\rm CAR} &-t_{\rm COT}(N_t-3) & 3\mu + H_c(N_t-3)
\end{pmatrix}
\end{equation}

When $N_t=3$, the diagonal elements become equal to $E_C$ at the charge degeneracy point $n_g=1,\, \mu=0$.
When the potential $\mu$ is tuned to achieve the degeneracy of the diagonal terms [Eq. \eqref{tuning}], the ground state of this sector results:
\begin{equation} \label{o1b}
\ket{3+}= \frac{1}{\sqrt{2}}\left[\cos\theta \ket{0100,N_t-1} + {\sf s}\cos\theta \ket{1000,N_t-1} + \sin\theta  \ket{0111,N_t-3} + {\sf s} \sin\theta \ket{1011,N_t-3} \right]\,
\end{equation} 
where ${\sf s}\equiv{\rm sign}\left[t_{\rm COT}(0) + t_{\rm COT}(2)\right]$. In this case, the parameter $\theta$ is defined as:
\begin{equation} \label{arg}
\theta=\frac{1}{2}{\rm Arg}\left[t_{\rm COT}(N_t-3) - t_{\rm COT}(N_t-1) +i 2\Delta_{\rm CAR}\right]\,.
\end{equation}
Eq. \eqref{o1} corresponds to the ground state \eqref{o1b} when $t_{\rm COT}(2)<t_{\rm COT}(0)<0$, thus ${\sf s}=-1$, and $\Delta_{\rm CAR}>0$ ($0<\alpha<\pi/4$ and $\mu_{\rm SC}<-2E_C\tilde{\Delta}/(\tilde{\Delta} + E_C)$). In this regime, considered in the main text, the energy of the ground states $\ket{3+}$ and $\ket{3-}$ results:
\begin{equation} \label{Eodd}
E^{(N_t=3)} = E_C\left[6-6n_g+n_g^2\right] 
-\bar{t}_{\rm COT}
- \sqrt{\left(\frac{t_{\rm COT}(0)-t_{\rm COT}(2)}{2}\right)^2 + \Delta^2_{\rm CAR}}\,,
\end{equation}
where we imposed the degeneracy condition \eqref{tuning}.

For generic $\mu$, the ground state maintains the form \eqref{o1b}, but the angle $\theta$ acquires a dependence on $\mu-2E_C(1-n_g)$, such that, in general, $\theta$ can be tuned through a suitable choice of $\mu$.
For general $N_t \neq 3$ and $\mu$, the diagonal elements in $H_{\rm eff}^{\text{odd}1}$ are indeed not degenerate, and the ground state energy acquires the form:
\begin{multline} \label{enodd}
E^{(N_t=2m+1)}= 2\mu + \frac{H_c(N_t-1)+H_c(N_t-3)}{2}  +{\sf s}\frac{\left( t_{\rm COT} (N_t-1) + t_{\rm COT} (N_t-3)\right)}{2} \\
-\sqrt{\frac{\left[2\mu + 4E_C\left(2+n_g-N_t \right) +{\sf s} \left( t_{\rm COT} (N_t-1) - t_{\rm COT} (N_t-3)\right)\right]^2}{4} + \Delta^2_{\rm CAR}}\,,
\end{multline}
with the sign {\sf s} now set to minimize the energy.
The ground state maintains the same form \eqref{o1b} with the angle $\theta$ suitably determined and the same dependence on the sign ${\sf s}$. More refined calculations can be performed by explicitly taking into account the dependence of $\mu$ from the cotunneling amplitudes in Eq. \eqref{hameff}.

We conclude by observing that in the special case in which $\mu_{SC}=0$, the cotunneling amplitudes in Eq. \eqref{cot} become opposite to each other. Therefore the sign ${\sf s}$ is not well defined. In this situation we obtain a two-fold degeneracy of the ground state of this symmetry sector when the degeneracy condition \eqref{tuning} is fulfilled.

\subsection{Odd sector $(N_t=2m+1,P_\Dnt=-)$}
This sector corresponds to the parities $P_\Upt=1,\,P_\Dnt=-1$ and is analogous to the previous under the exchange of the two nanowires $\Upt \leftrightarrow \Dnt$.
The basis is $\left\{\ket{0001,N_t-1},\ket{0010,N_t-1},\ket{1101,N_t-3},\ket{1110,N_t-3}\right\}$.

\begin{equation}
H_{\rm eff}^{\text{odd}-} = \begin{pmatrix} 
\mu +H_c(N_t-1) & -t_{\rm COT}(N_t-1) & -\Delta_{\rm CAR} & 0 \\
-t_{\rm COT}(N_t-1) & \mu +H_c(N_t-1) & 0 & -\Delta_{\rm CAR} \\
-\Delta_{\rm CAR} & 0 & 3\mu + H_c(N_t-3) & -t_{\rm COT}(N_t-3) \\
0 & -\Delta_{\rm CAR} &-t_{\rm COT}(N_t-3) & 3\mu + H_c(N_t-3)
\end{pmatrix}
\end{equation}
Analogously to the previous sector, for $N_t=3$ under the tuning condition \eqref{tuning}, the ground state of $(3,-)$ results:
\begin{equation}
\ket{3-} = \frac{1}{\sqrt{2}}\left[\cos\theta  \ket{0001,N_t-1} + {\sf s}\cos\theta\ket{0010,N_t-1} + \sin\theta \ket{1101,N_t-3} + {\sf s} \sin\theta \ket{1110,N_t-3} \right]\,,
\end{equation}
with $\theta$ is defined in Eq. \eqref{arg} and ${\sf s}$ is the sign of the cotunneling amplitude $t_{\rm COT}(N_{\rm max})$ with the largest modulus.

In the general case, the energy of the ground state matches Eq. \eqref{enodd}.

\subsection{Even sector $(N_t=2m,P_\Dnt=+)$} 
This sector corresponds to the parities $P_\Upt=1,\,P_\Dnt=1$.

Its basis is $\left\{\ket{0000,N_t},\ket{0011,N_t-2},\ket{1100,N_t-2},\ket{1111,N_t-4}\right\}$.

\begin{equation} \label{Heven1}
H_{\rm eff}^{\text{even}+} = \begin{pmatrix} 
H_c(N_t) & -\Delta_{\rm CAR} & -\Delta_{\rm CAR} & 0 \\
-\Delta_{\rm CAR} & 2\mu +H_c(N_t-2) & 0 & -\Delta_{\rm CAR} \\
-\Delta_{\rm CAR} & 0 & 2\mu + H_c(N_t-2) & -\Delta_{\rm CAR} \\
0 & -\Delta_{\rm CAR} &-\Delta_{\rm CAR} & 4\mu + H_c(N_t-4)
\end{pmatrix}
\end{equation}
This is the only sector in which three values of the charge of the superconducting island are involved. As a consequence, only three of the four diagonal entries can be tuned to be degenerate.
To study the role of the excited states in the definition of the Kondo problem, we represent the general form of the eigenstates of this Hamiltonian as:
\begin{equation} \label{e1g}
\ket{N_t,+,n} = A^{(n)}_{1,1}(N_t) \ket{0000,N_t} + A^{(n)}_{1,2}(N_t) \ket{0011,N_t-2}+ A^{(n)}_{1,3}(N_t) \ket{1100,N_t-2}+ A^{(n)}_{1,4}(N_t) \ket{1111,N_t-4}\,,\
\end{equation}
where we introduced the amplitudes $A^{(n)}_{c=1,j}$. Here the index $c=1$ labels the Hilbert sector $\left(P_\Up=1,P_\Dn=1\right)$, $j$ indicates the states in the basis of this sector and $n=1\ldots 4$ refers to the four eigenstates of this sector ordered by increasing eigenenergies. Therefore $\ket{4,+,1}$ corresponds to the ground state $\ket{4+}$ we analyzed in the main text. 

The sectors with total particle number $N_t=2,4$ present three equal diagonal terms when considering the tuning condition \eqref{tuning}.

We observe, in general, that the state $\left(\ket{0011,N_t-2} - \ket{1100,N_t-2} \right)/\sqrt{2}$ is an eigenstate of $H_{\rm eff}^{\text{even}1}$ with eigenenergy $2\mu +H_c(N_t-2)$. For $N_t=2,4$, this corresponds to the second energy state $n=2$. In particular, we have, $A_{1,3}^{(n)} = A_{1,2}^{(n)}$ for $n=1,3,4$ and $A_{1,3}^{(n)} = -A_{1,2}^{(n)} = 1/\sqrt{2}$ for $n=2$. 

For $N_t=4$, at the charge degeneracy point $n_g=1,\, \mu=0$, the diagonal entries of $H_{\rm eff}^{\text{even1}}$ become $9E_C,E_C,E_C,E_C$. In general, when $E_C \gg |\Delta_{CAR}|$ and $\mu < 2E_C$, the state $\ket{0000,4}$ acquires a considerably higher energy than the others and can be effectively removed from the low-energy basis. The Hamiltonian can then be approximated as:
\begin{equation}
H_{\rm eff}^{\text{even}+} (N_t=4)\xrightarrow{E_C \gg \Delta_{\rm CAR}} \begin{pmatrix} 
2\mu +H_c(2) & 0 & -\Delta_{\rm CAR} \\
 0 & 2\mu + H_c(2) & -\Delta_{\rm CAR} \\
 -\Delta_{\rm CAR} &-\Delta_{\rm CAR} & 4\mu + H_c(0)
\end{pmatrix}.
\end{equation}  
Eq. \eqref{even1_largeEc} corresponds to the ground state of this Hamiltonian. 
For $N_t=2$, the situation is analogous: in this case the state $\ket{1111,-2}$ acquires a higher energy and becomes decoupled for large $E_C$. In the limit of large $E_C$ the energies of the ground states $\ket{N_t=2,4,+}$ can thus be approximated as:
\begin{equation} \label{E1en}
E^{(N_t=2,4,+)} \xrightarrow{E_C \gg \Delta_{\rm CAR}} E_C\left[\left(N_t-2-n_g\right)^2 + 4\left(1-n_g\right)\right] 
-\sqrt{2}\Delta_{\rm CAR}\,.
\end{equation}

\subsection{Even sector $(N_t=2m,P_\Dnt=-)$}

This sector corresponds to $\left(P_\Upt=-1,P_\Dnt=-1\right)$ and its basis displays only states having the same charge $N_t-2$ in the superconducting island: $\left\{\ket{0101,N_t-2},\ket{0110,N_t-2},\ket{1001,N_t-2},\ket{1010,N_t-2}\right\}$.
\begin{equation}
H_{\rm eff}^{\text{even}-} = \begin{pmatrix} 
2\mu +H_c(N_t-2) & -t_{\rm COT}(N_t-2) & -t_{\rm COT}(N_t-2) & 0 \\
-t_{\rm COT}(N_t-2) & 2\mu +H_c(N_t-2) & 0 & -t_{\rm COT}(N_t-2) \\
-t_{\rm COT}(N_t-2) & 0 & 2\mu + H_c(N_t-2) & -t_{\rm COT}(N_t-2) \\
0 & -t_{\rm COT}(N_t-2) &-t_{\rm COT}(N_t-2) & 2\mu + H_c(N_t-2)
\end{pmatrix}
\end{equation}
Analogously to the previous sector, we introduce the following notation for the eigenstates of this sector:
\begin{equation} \label{e2g}
\ket{N_t,-,n} = A^{(n)}_{2,1} \ket{0101,N_t-2} + A^{(n)}_{2,2} \ket{0110,N_t-2} + A^{(n)}_{2,3} \ket{1001,N_t-2}+ A^{(n)}_{2,4} \ket{1010,N_t-2}\,.
\end{equation}
Given the structure of the Hamiltonian $H_{\rm eff}^{\text{even}2}$ all the coefficients $A^{(n)}_{2,j}$ have modulus $1/2$ and signs that depend on the sign of $t_{\rm COT}(N_t-2)$.

The ground state energy of this sector results:
\begin{equation} \label{Eeven2}
E^{(N_t=2m,-)}=E_C\left(N_t-2-n_g\right)^2 + 2\mu
-2\left|t_{\rm COT}(N_t-2)\right|\,.
\end{equation}

\section{The role of the even excited states in the definition of the Kondo problem} \label{app:exci}

In the main text we truncated the summation over the indices $n$ in Eqs. \eqref{Upsilon} and \eqref{Xi} to the even ground states ($n=1$) in Eqs. \eqref{even1_largeEc}, \eqref{even1_largeEc2} and \eqref{e2}. Here we consider the full summation in the definition of these operators to show how the topological Kondo point is modified by the inclusion of the excited states. In particular, we will obtain that the condition \eqref{tuningTK} to approach the topological Kondo point is slightly modified, and, in general, the excited states determine both the onset of additional marginal perturbations that do not affect its non-Fermi liquid behavior and additional contributions to the effective Zeeman terms. 

We consider the even states defined in Eqs. \eqref{e1g} and \eqref{e2g} for $N_t=4$ and $N_t=2$. Concerning the sector $(N_t=4,+)$, the matrix elements in Eq. \eqref{ddagm} are generalized to the following expressions:
\begin{align}
&\bra{4,+,n} d^\dag_\beta \ket{3+} = \frac{A_{1,3}^{(n)}(4)\cos\theta+A_{1,4}^{(n)}(4)\sin\theta}{\sqrt{2}}\delta_{\beta,1} - {\sf s} \frac{ A_{1,3}^{(n)}(4)\cos\theta+A_{1,4}^{(n)}(4)\sin\theta}{\sqrt{2}}\delta_{\beta,2} \equiv D_1^{(n)} \left(\delta_{\beta,1} - {\sf s}\delta_{\beta,2} \right) \,,\\
&\bra{4,+,n} d^\dag_\beta \ket{3-} = \frac{A_{1,2}^{(n)}(4)\cos\theta+A_{1,4}^{(n)}(4)\sin\theta}{\sqrt{2}}\delta_{\beta,3} -{\sf s}  \frac{A_{1,2}^{(n)}(4)\cos\theta+A_{1,4}^{(n)}(4)\sin\theta}{\sqrt{2}}\delta_{\beta,4}\equiv D_3^{(n)} \left( \delta_{\beta,3} - {\sf s} \delta_{\beta,4}\right) \,.
\end{align}
Importantly, given the structure of the Hamiltonian \eqref{Heven1} and the related coefficients $A_{1,j}^{(n)}$ we obtain that $D_1^{(n)} = D_3^{(n)}$ for $n=1,2,3$ and $D_1^{(2)} = -D_3^{(2)} = (\cos\theta)/2$.
Regarding the even sector 2, the matrix elements result:
\begin{align}
&\bra{4,-,n} d^\dag_\beta \ket{3+} = -\frac{\cos\theta}{\sqrt2}\left(A_{2,2}^{(n)}(4) + {\sf s} A_{2,4}^{(n)}(4)\right) \delta_{\beta,3} - \frac{\cos\theta}{\sqrt2} \left(A_{2,1}^{(n)}(4) + {\sf s} A_{2,3}^{(n)}(4)\right) \delta_{\beta,4} \equiv -C_3^{(n)}\delta_{\beta,3}-C_4^{(n)}\delta_{\beta,4}\,, \label{condi1}\\
&\bra{4,-,n} d^\dag_\beta \ket{3-} = \frac{\cos\theta}{\sqrt2}\left(A_{2,3}^{(n)}(4) + {\sf s}A_{2,4}^{(n)}(4)\right) \delta_{\beta,1} + \frac{\cos\theta}{\sqrt2} \left(A_{2,1}^{(n)}(4) + {\sf s} A_{2,2}^{(n)}(4)\right) \delta_{\beta,2} \equiv C_1^{(n)}\delta_{\beta,1}+C_2^{(n)}\delta_{\beta,2}\,. \label{condi2}
\end{align}
Based on these expressions, the terms entering the sum over $n$ to define $\Upsilon=\sum_n \Upsilon^{(n)}$ in Eq. \eqref{Upsilon} result:
\begin{align}
&\Upsilon_{(1,1)}^{(n)}=\frac{\left(D_1^{(n)}\right)^2}{\delta E_+^{(4,n)}} \left(\delta_{\alpha, 1} - {\sf s}\delta_{\alpha, 2}\right) \left(\delta_{\beta, 1} - {\sf s}\delta_{\beta, 2}\right) + \frac{1}{\delta E_-^{(4,n)}}\left(C_3^{(n)}\delta_{\alpha, 3} + C_4^{(n)}\delta_{\alpha, 4}\right) \left(C_3^{(n)}\delta_{\beta, 3} + C_4^{(n)}\delta_{\beta, 4}\right)\,,\\
&\Upsilon_{(1,2)}^{(n)}=\frac{D_1^{(n)}D_3^{(n)}}{\delta E_+^{(4,n)}} \left(\delta_{\alpha, 3} - {\sf s}\delta_{\alpha, 4}\right) \left(\delta_{\beta, 1} - {\sf s}\delta_{\beta, 2}\right) - \frac{1}{\delta E_-^{(4,n)}}\left(C_1^{(n)}\delta_{\alpha, 1} + C_2^{(n)}\delta_{\alpha, 2}\right) \left(C_3^{(n)}\delta_{\beta, 3} + C_4^{(n)}\delta_{\beta, 4}\right)\,,\\
&\Upsilon_{(2,1)}^{(n)}=\frac{D_1^{(n)}D_3^{(n)}}{\delta E_+^{(4,n)}} (\delta_{\alpha, 1} - {\sf s}\delta_{\alpha, 2}) (\delta_{\beta, 3} - {\sf s}\delta_{\beta, 4}) - \frac{1}{\delta E_-^{(4,n)}}\left(C_3^{(n)}\delta_{\alpha, 3} + C_4^{(n)}\delta_{\alpha, 4}\right)\left(C_1^{(n)}\delta_{\beta, 1} + C_2^{(n)}\delta_{\beta, 2}\right) \,,\\
&\Upsilon_{(2,2)}^{(n)}=\frac{\left(D_3^{(n)}\right)^2}{\delta E_+^{(4,n)}} \left(\delta_{\alpha, 3} - {\sf s}\delta_{\alpha, 4}\right) \left(\delta_{\beta, 3} - {\sf s}\delta_{\beta, 4}\right) + \frac{1}{\delta E_-^{(4,n)}}\left(C_1^{(n)}\delta_{\alpha, 1} + C_2^{(n)}\delta_{\alpha, 2}\right) \left(C_1^{(n)}\delta_{\beta, 1} + C_2^{(n)}\delta_{\beta, 2}\right)\,,
\end{align}
where the differences $\delta E_{c}^{(4,n)} = E^{(4,c)}_{n} - E^{(3)}$ match the denominators in the perturbative expression \eqref{Upsilon} and measure the energy gap between the $n^{\rm th}$ eigenstate of the even sector $(4,c)$ from the degenerate odd ground states.

To explicitly express the coefficients $C_{\beta}^{(n)}$, we introduce the sign ${\sf t}={\rm sign}\left[t_{\rm COT}(2)\right]$. If $\left|t_{\rm COT}(2)\right|>\left|t_{\rm COT}(0)\right|$, ${\sf s} = {\sf t}$ (this corresponds to $\mu_{SC}<-2E_C\tilde{\Delta}/(\tilde{\Delta}+ E_C)$ and $0<\alpha<\pi/4$ for $g_{\rm SC}=0$). Otherwise ${\sf s} = -{\sf t}$. In the following, we will address these two situations. 

As a function of the signs ${\sf s}$ and ${\sf t}$, the $C_{\beta}^{(n)}$ can be read in the table \ref{table:C}.

\begin{table}[h!]
\[
\begin{array}{|c|c|c|c|c|}
\hline
\displaystyle \frac{2\sqrt{2}C^{(n)}_\beta}{\cos \theta} & n=1 & n=2 & n=3 & n=4 \\[0.7em]
\hline
\beta=1 &{\sf s} + {\sf t} & -{\sf s} + {\sf t}&-{\sf s} - {\sf t}&{\sf s} - {\sf t} \\
\hline
\beta=2 &1+ {\sf s}{\sf t} & 1-{\sf s}{\sf t}&1+{\sf s}{\sf t}&1-{\sf s}{\sf t} \\
\hline
\beta=3 &{\sf s} + {\sf t} & -{\sf s} - {\sf t}&-{\sf s} + {\sf t}&{\sf s} - {\sf t} \\
\hline
\beta=4 &1+{\sf s} {\sf t} & 1+{\sf s}{\sf t}&1-{\sf s}{\sf t}&1-{\sf s}{\sf t} \\
\hline
\end{array}
\]
\caption{Table of the coefficients $\frac{2\sqrt{2}C^{(n)}_\beta}{\cos \theta}$ in Eqs. \eqref{condi1} and \eqref{condi2} as a function of ${\sf t}={\rm sign}\left[t_{\rm COT}(2)\right]$ and $\sf s={\rm sign}\left[t_{\rm COT}(2)+t_{\rm COT}(0) \right]$.}
\label{table:C}
\end{table}

Concerning the transitions mediated by states with $N_t=2$ electrons, we generalize the matrix elements in Eq. \eqref{dm} in the following form:
\begin{align}
&\bra{2,+,n} d_\alpha \ket{3+} = {\sf s}\frac{A_{1,1}^{(n)}(2)\cos\theta+A_{1,2}^{(n)}(2)\sin\theta}{\sqrt{2}}\delta_{\alpha,1} + \frac{ A_{1,1}^{(n)}(2)\cos\theta+A_{1,2}^{(n)}(2)\sin\theta}{\sqrt{2}}\delta_{\alpha,2} \equiv Q_1^{(n)} \left({\sf s}\delta_{\alpha,1} + \delta_{\alpha,2} \right) \,,\\
&\bra{2,+,n} d_\alpha \ket{3-} = {\sf s}\frac{A_{1,1}^{(n)}(2)\cos\theta+A_{1,3}^{(n)}(2)\sin\theta}{\sqrt{2}}\delta_{\alpha,3} +  \frac{A_{1,1}^{(n)}(2)\cos\theta+A_{1,3}^{(n)}(2)\sin\theta}{\sqrt{2}}\delta_{\alpha,4}\equiv Q_3^{(n)} \left(  {\sf s}\delta_{\alpha,3} + \delta_{\alpha,4}\right) \,, \\
&\bra{2,-,n} d_\alpha \ket{3+} = -\frac{\sin\theta}{\sqrt2}\left(A_{2,1}^{(n)}(2) + {\sf s} A_{2,3}^{(n)}(2)\right) \delta_{\alpha,3} + \frac{\sin\theta}{\sqrt2} \left(A_{2,2}^{(n)}(2) + {\sf s} A_{2,4}^{(n)}(2)\right) \delta_{\alpha,4} \equiv -P_3^{(n)}\delta_{\alpha,3}+P_4^{(n)}\delta_{\alpha,4}\,, \label{condi1b}\\
&\bra{2,-,n} d_\alpha \ket{3-} = \frac{\sin\theta}{\sqrt2}\left(A_{2,1}^{(n)}(2) + {\sf s}A_{2,2}^{(n)}(2)\right) \delta_{\alpha,1} - \frac{\sin\theta}{\sqrt2} \left(A_{2,3}^{(n)}(2) + {\sf s} A_{2,4}^{(n)}(2)\right) \delta_{\alpha,2} \equiv P_1^{(n)}\delta_{\alpha,1}-P_2^{(n)}\delta_{\alpha,2}\,. \label{condi2b}
\end{align}
The $Q^{(n)}_\alpha$ coefficients behave analogously to the $D^{(n)}_\beta$ coefficients, such that $Q^{(n)}_1=Q^{(n)}_3$ for $n=1,3,4$ and $Q^{(2)}_1=-Q^{(2)}_3$. The $P$ coefficients are instead summarized in table \ref{table:P} and depend on the sign of $t_{\rm COT}(0)$.

\begin{table}[h!]
\[
\begin{array}{|c|c|c|c|c|}
\hline
\displaystyle \frac{2\sqrt{2}P^{(n)}_\alpha}{\sin \theta} & n=1 & n=2 & n=3 & n=4 \\[0.7em]
\hline
\alpha=1 &1 + {\sf s} {\sf t} & 1+ {\sf s}{\sf t}&1-{\sf s}{\sf t}&1- {\sf s} {\sf t} \\
\hline
\alpha=2 &{\sf s} + {\sf t} & -{\sf s} - {\sf t}&-{\sf s} + {\sf t}&{\sf s} - {\sf t} \\
\hline
\alpha=3 &1+{\sf s} {\sf t} & 1-{\sf s}{\sf t}&1+{\sf s}{\sf t}&1-{\sf s}{\sf t} \\
\hline
\alpha=4 &{\sf s} + {\sf t} & -{\sf s} + {\sf t}&-{\sf s} - {\sf t}&{\sf s} - {\sf t}\\
\hline
\end{array}
\]
\caption{Table of the coefficients $\frac{2\sqrt{2}P^{(n)}_\alpha}{\sin \theta}$ in Eqs. \eqref{condi1b} and \eqref{condi2b} as a function of ${\sf t}={\rm sign}\left[t_{\rm COT}(0)\right]$ and ${\sf s}={\rm sign}\left[t_{\rm COT}(2)+t_{\rm COT}(0) \right]$.}
\label{table:P}
\end{table}

\subsection{Case $t_{\rm COT}(2) t_{\rm COT}(0)>0$}

Let us focus first on a regime of parameters such that both the cotunneling amplitudes $t_{\rm COT}(2)$ and $ t_{\rm COT}(0)$ share the same sign, therefore ${\sf s}={\sf t}$ for both the states at $N_t=2$ and $N_t=4$. This situation occurs when $|\mu_{SC}|>2E_C\tilde{\Delta} / (\tilde{\Delta} + E_C)$ (for $g_{SC}=0$).  

The regime with $t_{\rm COT}(2)<t_{\rm COT}(0)<0$ corresponds to the analysis in Sec. \ref{Kondo}. The situation is analogous for $0<t_{\rm COT}(2)<t_{\rm COT}(0)$. In these cases the coefficients $C^{(1)}_\beta$ and $P^{(1)}_\alpha$ do not vanish, such that the even ground states $\ket{N_t,-,1}$ play a major role in the definition of the matrices $\Upsilon$ and $\Xi$, thus on the Kondo problem. The excited states $\ket{N_t,-,4}$, instead, do not contribute to these operators since $C^{(4)}_\beta=P^{(4)}_\alpha=0$ for ${\sf t} = {\sf s}$.

In this regime, by considering the properties of the coefficients $D_\alpha^{(n)}$, we can rewrite the operators $\Upsilon$ in the form:
\begin{multline} \label{Upsilon1}
-\sum_n \Upsilon_{(1,1)}^{(n)}= \sum_n \left[\frac{D_1^{(n)}D_3^{(n)}}{\delta E_+^{(4,n)}}\right] \left(\delta_{\alpha, 1} - {\sf s}\delta_{\alpha, 2}\right) \left(\delta_{\beta, 1} - {\sf s}\delta_{\beta, 2}\right) + 
\frac{\cos^2\theta}{2\delta E_-^{(4,1)}}\left(\delta_{\alpha, 3} + {\sf s}\delta_{\alpha, 4}\right) \left(\delta_{\beta, 3} + {\sf s}\delta_{\beta, 4}\right) 
\\ + \frac{\cos^2\theta}{2\delta E_+^{(4,2)}} (\delta_{\alpha, 1} - {\sf s}\delta_{\alpha, 2}) (\delta_{\beta, 1} - {\sf s}\delta_{\beta, 2}) + \frac{\cos^2\theta}{2\delta E_-^{(4,2)}}\left(\delta_{\alpha, 3} - {\sf s}\delta_{\alpha, 4}\right) \left(\delta_{\beta, 3} - {\sf s}\delta_{\beta, 4}\right) \,,
\end{multline}
\begin{equation}
-\sum_n \Upsilon_{(1,2)}^{(n)}=\sum_n \left[\frac{D_1^{(n)}D_3^{(n)}}{\delta E_+^{(4,n)}} \right] \left(\delta_{\alpha, 3} - {\sf s}\delta_{\alpha, 4}\right) \left(\delta_{\beta, 1} - {\sf s}\delta_{\beta, 2}\right) - \frac{\cos^2\theta}{2\delta E_-^{(4,1)}}\left(\delta_{\alpha, 1} + {\sf s}\delta_{\alpha, 2}\right) \left(\delta_{\beta, 3} +{\sf s} \delta_{\beta, 4}\right) \,,
\end{equation}
\begin{equation}
-\sum_n\Upsilon_{(2,1)}^{(n)}=\sum_n \left[\frac{D_1^{(n)}D_3^{(n)}}{\delta E_+^{(4,n)}}\right] (\delta_{\alpha, 1} - {\sf s}\delta_{\alpha, 2}) (\delta_{\beta, 3} - {\sf s}\delta_{\beta, 4}) - \frac{\cos^2\theta}{2\delta E_-^{(4,1)}}\left(\delta_{\alpha, 3} + {\sf s}\delta_{\alpha, 4}\right)\left(\delta_{\beta, 1} + {\sf s}\delta_{\beta, 2}\right) \,,
\end{equation}
\begin{multline} \label{Upsilon4}
-\sum_n \Upsilon_{(2,2)}^{(n)}= \sum_n \left[\frac{D_1^{(n)}D_3^{(n)}}{\delta E_+^{(4,n)}}\right] \left(\delta_{\alpha, 3} - {\sf s}\delta_{\alpha, 4}\right) \left(\delta_{\beta, 3} - {\sf s}\delta_{\beta, 4}\right) + 
\frac{\cos^2\theta}{2\delta E_-^{(4,1)}}\left(\delta_{\alpha, 1} + {\sf s}\delta_{\alpha, 2}\right) \left(\delta_{\beta, 1} 
+ {\sf s}\delta_{\beta, 2}\right) \\
+ \frac{\cos^2\theta}{2\delta E_+^{(4,2)}} (\delta_{\alpha, 3} - {\sf s}\delta_{\alpha, 4}) (\delta_{\beta, 3} - {\sf s}\delta_{\beta, 4}) + \frac{\cos^2\theta}{2\delta E_-^{(4,2)}}\left(\delta_{\alpha, 1} - {\sf s}\delta_{\alpha, 2}\right) \left(\delta_{\beta, 1} - {\sf s}\delta_{\beta, 2}\right)\,.
\end{multline}
The operators $\Xi$ result instead:
\begin{multline} \label{Xi1}
-\sum_n \Xi_{(1,1)}^{(n)}= \sum_n \left[\frac{Q_1^{(n)}Q_3^{(n)}}{\delta E_+^{(2,n)}}\right] \left(\delta_{\alpha, 1} + {\sf s}\delta_{\alpha, 2}\right) \left(\delta_{\beta, 1} + {\sf s}\delta_{\beta, 2}\right) + 
\frac{\sin^2\theta}{2\delta E_-^{(4,1)}}\left(\delta_{\alpha, 3} - {\sf s}\delta_{\alpha, 4}\right) \left(\delta_{\beta, 3} - {\sf s}\delta_{\beta, 4}\right) 
\\ + \frac{\sin^2\theta}{2\delta E_+^{(2,2)}} (\delta_{\alpha, 1} + {\sf s}\delta_{\alpha, 2}) (\delta_{\beta, 1} + {\sf s}\delta_{\beta, 2}) + \frac{\sin^2\theta}{2\delta E_-^{(2,2)}}\left(\delta_{\alpha, 3} + {\sf s}\delta_{\alpha, 4}\right) \left(\delta_{\beta, 3} + {\sf s}\delta_{\beta, 4}\right) \,,
\end{multline}
\begin{equation}
-\sum_n \Xi_{(1,2)}^{(n)}=\sum_n \left[\frac{Q_1^{(n)}Q_3^{(n)}}{\delta E_+^{(2,n)}} \right] \left(\delta_{\alpha, 3} + {\sf s}\delta_{\alpha, 4}\right) \left(\delta_{\beta, 1} + {\sf s}\delta_{\beta, 2}\right) - \frac{\sin^2\theta}{2\delta E_-^{(2,1)}}\left(\delta_{\alpha, 1} - {\sf s}\delta_{\alpha, 2}\right) \left(\delta_{\beta, 3} -{\sf s} \delta_{\beta, 4}\right) \,,
\end{equation}
\begin{equation}
-\sum_n\Xi_{(2,1)}^{(n)}=\sum_n \left[\frac{Q_1^{(n)}Q_3^{(n)}}{\delta E_+^{(2,n)}}\right] (\delta_{\alpha, 1} + {\sf s}\delta_{\alpha, 2}) (\delta_{\beta, 3} + {\sf s}\delta_{\beta, 4}) - \frac{\sin^2\theta}{2\delta E_-^{(2,1)}}\left(\delta_{\alpha, 3} - {\sf s}\delta_{\alpha, 4}\right)\left(\delta_{\beta, 1} - {\sf s}\delta_{\beta, 2}\right) \,,
\end{equation}
\begin{multline} \label{Xi4}
-\sum_n \Xi_{(2,2)}^{(n)}= \sum_n \left[\frac{Q_1^{(n)}Q_3^{(n)}}{\delta E_+^{(2,n)}}\right] \left(\delta_{\alpha, 3} + {\sf s}\delta_{\alpha, 4}\right) \left(\delta_{\beta, 3} + {\sf s}\delta_{\beta, 4}\right) + 
\frac{\sin^2\theta}{2\delta E_-^{(2,1)}}\left(\delta_{\alpha, 1} - {\sf s}\delta_{\alpha, 2}\right) \left(\delta_{\beta, 1} 
- {\sf s}\delta_{\beta, 2}\right) \\
+ \frac{\sin^2\theta}{2\delta E_+^{(2,2)}} (\delta_{\alpha, 3} + {\sf s}\delta_{\alpha, 4}) (\delta_{\beta, 3} + {\sf s}\delta_{\beta, 4}) + \frac{\sin^2\theta}{2\delta E_-^{(2,2)}}\left(\delta_{\alpha, 1} + {\sf s}\delta_{\alpha, 2}\right) \left(\delta_{\beta, 1} + {\sf s}\delta_{\beta, 2}\right)\,.
\end{multline}

When considering the matrices $\Sigma= \Upsilon-\Xi$, a possible fine-tuning to retrieve a perturbed topological Kondo point is the following:
\begin{equation} \label{TKtuning}
\sum_n \left[\frac{D_1^{(n)}D_3^{(n)}}{\delta E_+^{(4,n)}} + \frac{Q_1^{(n)}Q_3^{(n)}}{\delta E_+^{(2,n)}}\right]= \frac{\cos^2\theta}{2\delta E_-^{(4,1)}} + \frac{\sin^2\theta}{2\delta E_-^{(2,1)}}\,.
\end{equation}
This tuning generalizes the condition \eqref{tuningTK} discussed in Sec. \ref{PMTK} by including additional contributions of the excited states to the left-hand side of the equation. The overall effect of the excited states amounts to a correction of a few percent in the definition of the $\Sigma$ matrix elements, since the terms with $n=2$ and $n=3,4$ contribute with opposite signs. A comparison between the term $A_+$ (Eq. \eqref{Aplus}) which represents the $n=1$ contribution, and the full expression given by the left hand side of Eq. \eqref{TKtuning} is displayed in Fig.~\ref{fig:exci}. The optimal topological Kondo point is met when the latter crosses $B_+$ (Eq. \eqref{Bplus}), which constitutes the right hand side of Eq. \eqref{TKtuning}.
Such constraint allows us to obtain a topological Kondo Hamiltonian in the form \eqref{hamkondo2} with the addition of the terms in the second lines of Eqs. \eqref{Upsilon1}, \eqref{Upsilon4}, \eqref{Xi1} and \eqref{Xi4}. 

\begin{figure}[t]
\includegraphics[width=10cm]{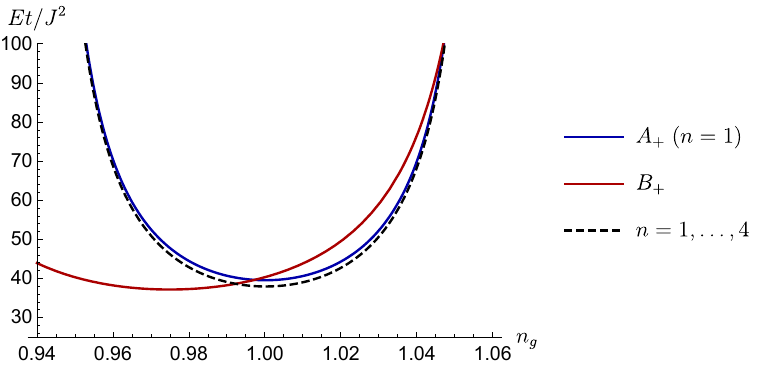}
\caption{Comparison between the contribution $A_+$ of the even ground states $\ket{N_t,+}$ (blue line) and the full expression including also the excited states (black dashed line) in the definition of the $\Sigma$ matrices, as a function of $n_g$. The topological Kondo conditions in Eqs. \eqref{tuningTK} and \eqref{TKtuning} correspond to the crossings of these lines with the red line that depicts $B_+$. The chosen parameters are $\Delta=4t$, $\mu_{SC}=-5t$, $E_C=0.15t$ and $\alpha=\pi/6$. These matrix elements are plotted in units of $J^2/t$.}
\label{fig:exci}
\end{figure}

Let us consider these additional perturbations. For $\alpha=\beta$, these terms can be corrected through a suitable shift of the potentials in the leads, and do not qualitatively affect the Kondo problem apart for a correction of the Zeeman term $\sigma_3$ entering Eq. \eqref{hamkondof} for the 3-lead setups.
For $\alpha=1$ and $\beta=2$ or viceversa, instead, these terms provide a potential scattering contribution ($l^\dag_2 l_1 \sigma_0 + {\rm H.c.}$), analogous to the ones discussed in Sec. \ref{sec:RG}, which does not renormalize and does not affect the non-Fermi liquid behavior of the topological Kondo point \cite{Beri2013,Galpin2014}. Symmetric terms are generated in the four-lead devices for $\alpha=3$ and $\beta=4$ or viceversa.

\subsection{Case $t_{\rm COT}(2) t_{\rm COT}(0)<0$}

If $\left|\mu_{\rm SC} \right| < 2E_C\tilde{\Delta} / (\tilde{\Delta} + E_C)$ (for $g_{SC}=0$), the cotunneling amplitudes $t_{\rm COT}(2)$ and $t_{\rm COT}(0)$ acquire opposite signs. Depending on whether $|t_{\rm COT}(0)|\gtrless|t_{\rm COT}(2)|$ (for $\mu_{\rm SC} \gtrless 0$ respectively), either the state $\ket{4,-,1}$ or $\ket{2,-,1}$ do not contribute to the definition of the $\Upsilon$ or $\Xi$ matrices respectively. This is because the related coefficients $C_\beta^{(1)}$ or $P_\alpha^{(1)}$ vanish. The role of the missing $n=1$ state is replaced by the higher energy state $n=4$, whose contribution, however, is reduced by the larger denominator $\delta E_-^{(N_t,4)}$. In this situation it is easy to rederive the related $\Upsilon$ and $\Xi$ matrices.
The corresponding fine-tuning conditions to approach the topological Kondo point result:
\begin{align}
& \sum_n \left[\frac{D_1^{(n)}D_3^{(n)}}{\delta E_+^{(4,n)}} + \frac{Q_1^{(n)}Q_3^{(n)}}{\delta E_+^{(2,n)}}\right]= \frac{\cos^2\theta}{2\delta E_-^{(4,4)}} + \frac{\sin^2\theta}{2\delta E_-^{(2,1)}}\,, \quad \text{for }  |t_{\rm COT}(0)|>|t_{\rm COT}(2)|\,,\\
& \sum_n \left[\frac{D_1^{(n)}D_3^{(n)}}{\delta E_+^{(4,n)}} + \frac{Q_1^{(n)}Q_3^{(n)}}{\delta E_+^{(2,n)}}\right]= \frac{\cos^2\theta}{2\delta E_-^{(4,1)}} + \frac{\sin^2\theta}{2\delta E_-^{(2,4)}}\,, \quad \text{for }  |t_{\rm COT}(0)|<|t_{\rm COT}(2)|\,.
\end{align}
\end{widetext}

Our analysis of the Kondo problems stemming from the poor man's tetron holds as long as the global ground states are in the odd $N_t=3$ sector and the system does not display the additional degeneracy at $\mu_{SC}=0$. If, however,  $t_{\rm COT}(2)=-t_{\rm COT}(0)$, the odd $N_t=3$ sectors acquire an additional degeneracy when the tuning \eqref{tuning} is fulfilled. In this situation the study of the related Kondo physics would require to treat the system as an effective spin 3/2 impurity rather than a spin 1/2 impurity. Our perturbative results, however, suggest that the global ground state of the poor man's tetron does not belong to the degenerate $N_t=3$ sectors for $\mu_{SC}=\mu=0$ and $n_g=1$, such that a four-fold degeneracy of the ground state is not realized, at least in the small $t$ regime.

\bibliography{MPStransport}
\end{document}